\documentclass[pra,superscriptaddress]{revtex4}

\usepackage{amsmath}
\usepackage{graphicx}
\usepackage{dsfont}
\usepackage{epstopdf}
\usepackage{amsfonts}
\usepackage{color}
\usepackage{hyperref}
\usepackage{bm}

\definecolor{darkblue}{rgb}{0,0,0.75}
\definecolor{darkred}{rgb}{0.5,0,0}
\definecolor{dg}{rgb}{0,0.3,0}

\hypersetup{
 pdfborder={0,0,0},
 colorlinks=true,
 linkcolor=darkblue,
 citecolor = darkblue
} 

\begin{document}

\title{Propagation of radiation pulses through gas-plasma mixtures}

\author{Karl-Peter Marzlin}
\affiliation{Department of Physics, St. Francis Xavier University,
  Antigonish, Nova Scotia B2G 2W5, Canada}
\affiliation{Department of Physics and Atmospheric Science,
        Dalhousie University, Halifax, Nova Scotia B3H 4R2, Canada}

\author{Anuraj Panwar}
\affiliation{Department of Physics, POSTECH, Hyoja-Dong San 31, 
KyungBuk, Pohang 790-784, South Korea}

\author{M. Shajahan G. Razul}
\affiliation{Department of Physics, St. Francis Xavier University,
  Antigonish, Nova Scotia B2G 2W5, Canada} 

\author{Barry C.~Sanders}
\affiliation{Institute for Quantum Science and Technology,
        University of Calgary, Calgary, Alberta T2N 1N4, Canada}
\affiliation{%
    Program in Quantum Information Science,
    Canadian Institute for Advanced Research,
    Toronto, Ontario M5G 1Z8, Canada%
    }

\begin{abstract}
We determine the linear optical susceptibility of a radiation pulse 
propagating through a mixture of a gas of atoms or molecules and a plasma. 
For a specific range of radiation and plasma frequencies,
resonant generation of volume plasmons significantly
amplifies the radiation intensity.
The conditions for resonant
amplification are derived from the dispersion relations in the
mixture, and the amplification is demonstrated in a numerical
simulation of pulse propagation.
\end{abstract}



\maketitle

\section{Introduction}
In applications of 
electromagnetic fields, radiation sometimes
consists of a few photons only
\cite{lukin:NatPhys2007} or interacts with a single molecule
\cite{Science275-1102}.
As source and/or signal in such
experiments are weak, it is necessary to amplify the interaction
between matter and radiation. This has already been accomplished in a
variety of ways
\cite{Schmidt96,Hau99,Kang03,wang:063901,PhysRevLett.113.173601},
but a method that would target a specific spatial area in an adjustable
way would be useful for designing experiments. 

In recent years, a number of studies have addressed
propagation of electromagnetic waves in an
ultra-cold neutral plasma.
Radio frequency fields have been used to study
collective plasma electron oscillations and plasma
expansion in ultra-cold neutral plasmas~\cite{Kulin2000, Fletcher2006}.
In the optical regime these experiments are complemented
by absorption imaging methods to determine the
ion velocity distribution~\cite{Simien2004,Killian2005}. 
If radiation propagates through a gas of Rydberg atoms
\cite{Mendonca2009}
wave instabilities can occur due to
energy transfer between excited Rydberg states and plasma electrons,
and electrostatic waves \cite{Shukla2010}.
Lu et al.~\cite{Ronghua2011}  suggested microwaves
as a tool to measure the recombination rate of electrons and ions in 
ultra-cold neutral plasmas. 
Mendon\c{c}a et al.~\cite{Mendonca2010} 
predicted the generation of quasi-stationary magnetic fields
by high-intensity radiation in Rydberg plasmas.

Radiation propagation through gas-plasma mixtures has also been 
studied for warm or hot plasmas. Plasma-induced optical sidebands
around forbidden transitions
in atomic spectra have been predicted \cite{Baranger1961} 
and demonstrated experimentally \cite{Kunze1968,Cooper1969} long ago.
During the past two decades, the focus has been on nonlinear effects
induced by high-intensity radiation. 
Among the most interesting nonlinear phenomena in mixtures
are relativistic guiding and self-modulation of short laser pulses 
\cite{Sprangle1990a,Sprangle1990b,Sprangle1992,Esarey1994},
harmonic generation and refraction
\cite{Leemans1992a,Mackinnon1996,Esarey1997}, as well as 
de-focusing \cite{Sprangle1996} and dispersion \cite{Wu2003}. 
These effects are relevant to the construction of 
laser-driven plasma-based electron accelerators \cite{RevModPhys.81.1229}.
Hu et al.~\cite{Hu2005} predicted self-generation of a
quasi-static magnetic field for short, intense laser pulses.

In this paper we determine under which conditions
volume plasmons \cite{PhysRevLett.102.156802} can be used
to amplify a radiation pulse. 
Volume plasmons are electron density waves inside a plasma,
which are related to surface plasmons at the interface between
a metal and a dielectric. Our work is guided by the idea that
the well-known method to amplify radiation using surface plasmons 
\cite{:/content/aip/journal/jcp/136/14/10.1063/1.3698292}
may be extended to volume plasmons by employing a 
mixture of an atomic or molecular gas and plasma.
The advantage of such a scheme would be that the amplification
would not require the close (tens of nm) proximity of a metal
surface. Furthermore, it would be possible to control the
location where radiation is amplified, so that a specific set
of molecules in a sample could be targeted.
While we are mainly interested in the propagation of a 
controlled radiation pulse, our findings may also be of relevance
for radiation propagation through the ionosphere or ionized interstellar media.

\begin{figure}
\begin{center}
\includegraphics[width=9cm]{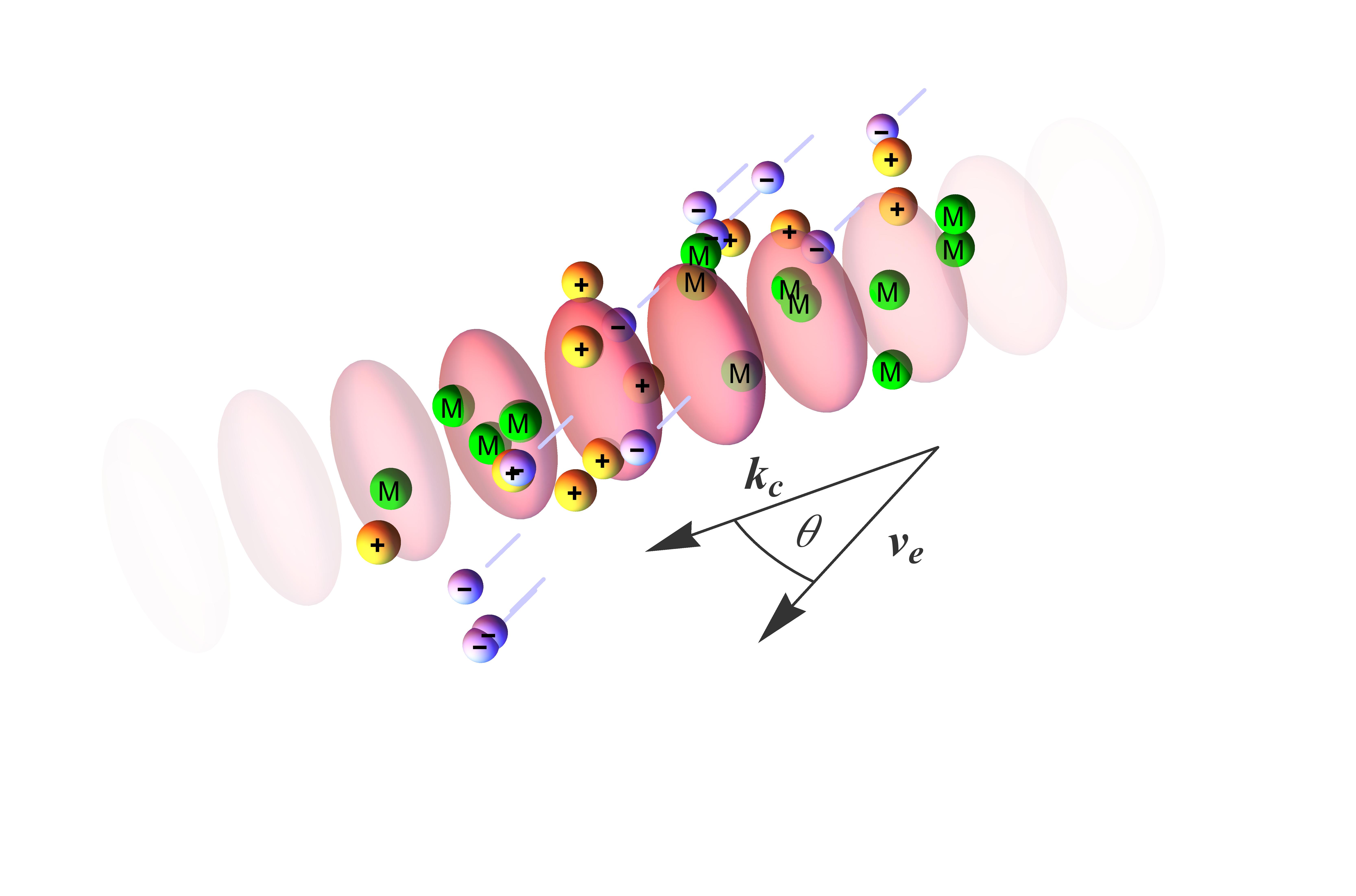}
\caption{\label{fig:sketch}
A radiation pulse (whose intensity maxima are
  represented by ellipses) with central wavevector $\bm{k}_c$
propagates through a mixture of a gas of 
 atoms or molecules and a plasma. Molecules or atoms (M) and positive ions (+) are virtually immobile
during the pulse propagation time. Plasma electrons (-) move with mean
velocity $\bm{v}_\text{e}$ at an angle
$\theta$ relative to the axis of the radiation pulse.}
\end{center}
\end{figure}

We derive an expression for the
linear index of refraction for radiation propagating through the mixture.
We show that a strong amplification may occur if 
volume plasmons are generated and demonstrate this effect by a
numerical simulation of pulse propagation through the medium.
Amplification is resonantly enhanced for specific radiation frequencies $\omega$
and angles $\theta$ between the
radiation pulse and the velocity $\bm{v}_\text{e}$ of plasma electrons (see
Fig.~\ref{fig:sketch}). 

The paper is organized as follows. In
Sec.~\ref{sec:physDescr} we discuss the general physical features
of radiation-gas-plasma systems. 
Theoretical methods and results are summarized
in Sec.~\ref{sec:theoResults}, followed by a detailed
discussion of optical dispersion relations and field amplitudes 
in Sec.~\ref{sec:dispersion}. Numerical results for pulse propagation
through a mixture are presented in Sec.~\ref{sec:pulsePropagation}.
Several appendices contain the details of our calculations.

\section{Coupled radiation-atom-plasma systems}\label{sec:physDescr}
The physical case that we study is sketched in
Fig.~\ref{fig:sketch}.
A radiation pulse propagates through a mixture of an atomic gas and a
plasma. For simplicity we refer to the gas particles as atoms,
even though the gas may be composed of atoms or molecules.
The plasma electrons move with (mean) velocity
$\bm{v}_\text{e}$.
To derive the linear optical susceptibility of this system
and to simulate the propagation of a radiation pulse, we 
solve the coupled equations of
motion for the radiation-atom-plasma system.
In this section we
describe the general features of these equations. Full details are given 
in App.~\ref{sec:dynnamicalEquations}.

\subsection{Plasma component}
The plasma is modelled as a classical gas of ions and electrons, 
both of which may have a thermal 
distribution. For sufficiently short radiation pulses, the relatively slow motion 
of the ions may be neglected. The plasma dynamics can then be
described by a phase-space distribution $f(\bm{r}, \bm{ p}, t)$ of electrons. In
the kinetic theory of electron gases, this distribution obeys a
variant of the Boltzmann equation, which is called the Vlasov equation
\cite{SitenkoMalnev1995}. We use a relativistic generalization 
of the Vlasov equation \cite{Liboff.R.L2003}.

The Vlasov equation is non-linear in the dynamical fields,
but for a weak radiation field it can be linearized in 
the deviation \cite{GoldstonRutherford}
\begin{align} 
  \delta f(\bm{ r}, \bm{ p},t) = f(\bm{ r}, \bm{ p}, t) -f_0(\bm{ p})  ,
\label{eq:eqLinDist}\end{align} 
of $f(\bm{ r}, \bm{ p}, t) $ from the initial 
distribution $ f_0(\bm{ p}) $, which is assumed to be spatially
homogeneous.
Our methods can be extended to spatially
   inhomogeneous initial distributions $ f_0(\bm{r},\bm{ p}) $,
but the required numerical resources are much larger than in the
homogeneous case.
The linearized relativistic Vlasov equation then takes the form
\begin{align} 
  0 &=   \partial_t \delta f + \frac{1}{m\gamma}\bm{ p} \cdot \nabla_{\bm{r}} \delta f - q
  \left ( \bm{E} +  \frac{ 1}{m\gamma}\bm{ p}\times \bm{ B} \right ) \cdot \nabla_{\bm{p}} f_0,
\label{eq:vlasov}\end{align} 
where $q$ and $m$ denote fundamental charge and 
electron mass, respectively. The last term describes the influence of
the electromagnetic Lorentz force on the electrons, with
$\gamma^2 = 1+ (p/mc)^2$ the Lorentz factor.

\subsection{Atomic gas}
The detailed theory that is presented in
App.~\ref{sec:dynnamicalEquations} 
includes a full quantum field 
description of atomic observables, so that the methods developed here
can be applied to molecular spectroscopy \cite{doi:10.1021/jp800167v}
and quantum information
\cite{Ferguson2002,Schroder2009,1367-2630-11-5-055022}. 
However, in this paper we focus on linear optical properties of 
atom-plasma mixtures, which are essentially of classical nature.
We therefore describe the radiation field and plasma
electrons using classical fields;
only the internal  (electronic)
dynamics of atoms is treated quantum mechanically.
The quantum nature of the atomic center-of-mass dynamics is only 
relevant at temperatures well below 1 mK 
\cite{TannoudjiLesHouches1990}, which is hard to realize even
in ultracold atom-plasma mixtures.
At temperatures above 1 mK, the center-of-mass motion can be
included via Doppler broadening of spectral lines
\cite{AtomPhotonInteractions}. 

Within this approximation, the atomic degrees of freedom can be
described by a coherence field $\sigma_{ij}(\bm{ r},t)$, which probes
to what degree the atoms are prepared in a superposition
of internal electronic states $|i \rangle $ and $|j \rangle $.
In this work we focus on near-resonant electromagnetic fields,
  so that only the two atomic levels that are resonantly coupled need
  to be taken into account. We therefore can consider
two-level atoms 
with ground state $|g \rangle $ and excited state $|e \rangle $.
The polarization field is then related to the electromagnetic
polarization field of the atomic gas through
\begin{equation} 
  \bm{ P} (\bm{ r},t)=   {\cal N}_\text{{\tiny A}}   \bm{ d}_{eg}
  \sigma_{eg}(\bm{ r},t)  +\text{ c.c.}\; ,
\label{eq:atomPolariz}\end{equation} 
with the atomic density ${\cal N}_\text{{\tiny A}}$. Here
$\bm{ d}_{eg} = \langle e | \hat{\bm{d}} | g \rangle $ is the atomic
dipole moment, with $\hat{\bm{d}}$  the dipole operator.
Hence, one may think of the coherence field as a re-scaled, complex
form of the polarization field. Its equation of motion
is derived in App.~\ref{sec:dynnamicalEquations}  
and given by 
\begin{align} 
  \partial_t \sigma_{eg}(\bm{ r},t) &=
  \left  (-i \omega_0 -\frac{ \Gamma}{2}\right ) \sigma_{eg}(\bm{ r},t)  +\frac{i}{\hbar\varepsilon_0}   
  \bm{D}_\perp(\bm{ r},t) \cdot  \bm{ d}_{eg}^{\, *} ,
\label{eq:atomdynamics}\end{align} 
where $\omega_0$ denotes the atomic resonance frequency.
The atomic spectral linewidth $\Gamma$
includes the effects of spontaneous emission as
well as the influence of the environment, such as Doppler and
collisional broadening.
The field
$\bm{D}_\perp(\bm{ r},t)$ represents the transverse part 
($\nabla\cdot \bm{D}_\perp =0$) of the electric displacement field
$\bm{D}$. 

\subsection{Radiation Field}
The radiation field is described using the
macroscopic Maxwell equations
\begin{align} 
  \nabla \cdot \bm{D} &= \rho ,
\label{eq:Maxw1}\\
   \nabla \cdot \bm{B} &= 0,
\\
  \nabla \times \bm{E} &= - \partial_t \bm{B},
\\
     \nabla \times \bm{B} &= \mu_0 \left (\bm{J}  + \partial_t \bm{D}
     \right ),
\label{eq:Maxw4}\end{align} 
with free charge density $\rho$  and free current density $\bm{J}$
provided by the plasma particles, and bound charges corresponding
to the atoms. We assume that the magnetization field of the 
atom-plasma mixture is negligible.
The atoms affect radiation through their polarization
field $\bm{ P}$, which appears in the material equation $ \bm{ D} = \varepsilon_0 \bm{ E} + \bm{ P}$.
To describe the interaction between atoms and 
a plane electromagnetic wave with wavevector $\bm{k}$
we introduce a set of basis vectors
\begin{figure}
\begin{center}
\includegraphics[width=5cm]{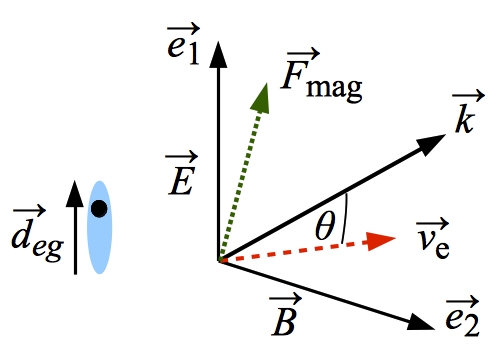}
\caption{\label{fig:forceDiagram}
A radiation pulse with wavevector $\bm{ k}$ is polarized in the direction
of the dipole moment $\bm{ d}_{eg}$ of the atoms. Depending on the
direction of the plasma electron velocity $\bm{ v}_\text{e}$, the
magnetic force $\bm{ F}_\text{mag}$ on the electrons can possess a
component parallel to the wavevector.} 
\end{center}
\end{figure}
\begin{align} 
    \hat{\bm{k}} := \frac{ \bm{ k}}{|\bm{ k}|}
  \; , \; 
  \bm{ e}_{{ 1}} (\hat{\bm{k}}) = \frac{  \bm{
      d}_{\perp}(\hat{\bm{k}}) 
  }{\left |\bm{ d}_{\perp }(\hat{\bm{k}}) \right |} 
\; , \;
 \bm{ e}_{{ 2}} (\hat{\bm{k}})&= 
  \bm{ e}_{{ 1}}^* (\hat{\bm{k}}) \times \hat{\bm{k}},
\label{def:e1}
\end{align} 
where  $\hat{\bm{k}}$ is a unit vector that is parallel to the wavevector
and
\begin{equation} 
   \bm{d}_{\perp} (\hat{\bm{k}}) := 
   \bm{ d}_{eg} -  
  \left (\hat{\bm{k}}\cdot  \bm{ d}_{eg}\right ) 
   \hat{\bm{k}} \; ,
\label{eq:dPerpDef}\end{equation} 
the part of $\bm{ d}_{eg}$ that is perpendicular to $\hat{\bm{k}}$.
For radiation polarized along $\bm{ e}_{{ 1}}
$, and the wavevector is perpendicular to $\bm{ d}_{eg}$,
the interaction between atoms and radiation is maximized.
This situation is depicted in Fig.~\ref{fig:forceDiagram}.
If the polarization points along  $\bm{ e}_{{ 2}}$, the
atoms appear transparent to the radiation pulse.

\subsection{Volume plasmons}
The interaction between plasma electrons and radiation 
enters into the Vlasov equation (\ref{eq:vlasov}) through
the Lorentz force $\bm{ F}=-q \bm{ E} -q \bm{ v}_\text{e} \times
\bm{ B}$. 
In free space, electric
and magnetic field amplitude of a radiation pulse are related through
$\bm{ B} = \hat{\bm{k}}\times \bm{ E}/c$. For this reason, the magnetic
force $\bm{ F}_\text{mag} = -q \bm{ v}_\text{e} \times
\bm{ B}$ is suppressed unless the electrons travel at relativistic speed.
Hence, at low velocities, the electrons are only accelerated by the
electric field in a direction perpendicular to the radiation pulse.

When $v_\text{e}$ is comparable to the speed of
light $c$, the force $\bm{ F}_\text{mag} $ cannot be neglected
anymore. The fact that  the magnetic force is always perpendicular to both
$\bm{ B}$ and $\bm{ v}_\text{e}$ enables it to induce longitudinal
modulations of the electron density, i.e., volume plasmons
\cite{PhysRevLett.102.156802}, as long as
the electron velocity is not exactly parallel to the
wavevector. 
In Eq.~(\ref{eq:vlasov}), these modulations are generated by the term
proportional to $\nabla_{\bm{p}}f_0$, which for a narrow momentum
distribution is non-zero only for momenta $\bm{p}$ that are close to
the initial mean momentum $\bm{p}_e$ of the electrons. This term acts
as a source term for spatial variations, which are generated through
the term involving  $\nabla_{\bm{r}}\delta f$.

Plasmon generation can lead to dramatic changes
of radiation dynamics, including amplification of light intensities
near a metal surface by several orders of magnitude
\cite{KneippSERS2006}. The discussion above suggests that the 
scenario depicted in
Fig.~\ref{fig:forceDiagram} to generate volume plasmons inside an
atom-plasma mixture will be most promising for light amplification.
The radiation polarization is parallel to the
atomic dipole moment, and plasma electrons move at relativistic
speed with a velocity component along $\bm{ e}_{{ 1}}$.
The radiation pulse then interacts strongly with the atoms and can induce
volume plasmons. If the electrons move in the direction of $\bm{
  e}_{{ 2}}$ instead, they could still form volume plasmons,
but only for radiation that does not interact with atoms.

Plasma electrons with relativistic speed are not only interesting 
because of the increase of $\bm{ F}_\text{mag}$, but also because
they may interact resonantly with radiation. Generally, charged particles
are most effectively accelerated if they move at the same speed
as the phase front of an electromagnetic wave. In free space this is
impossible, but in the presence of a dielectric medium with refractive
index $n$, electrons are strongly interacting with radiation if their
longitudinal velocity is $v_\text{e} = c/n$
\cite{RevModPhys.81.1229}. 

In our case, this medium is formed
by the atomic gas. Below we show that the absorption of radiation
by atoms modifies the resonance condition  $v_\text{e} = c/n$.
A main result of our paper is to show that
resonances can still occur, but only at specific optical frequencies,
and for a specific range of electron densities.

\section{Theoretical Results}\label{sec:theoResults}
Solving the dynamics of radiation in an atom-plasma mixture
is a lengthy and rather tedious process. In this section,
we give a short summary of our methods and present the
general results. The details of the derivation are presented in
App.~\ref{app:solution}.

The atom-plasma mixture is
initially homogeneous, with the  atoms prepared in their ground state.
At time $t=0$, a weak radiation pulse with initial electric field amplitude
$\bm{ E}_0( \bm{ r})$ is switched on.  
The dynamical equations of our system,
Eqs.~(\ref{eq:vlasov}) for the plasmon dynamics,
(\ref{eq:atomdynamics}) for the atomic evolution, and the
macroscopic Maxwell equations (\ref{eq:Maxw1})-(\ref{eq:Maxw4}) 
for radiation propagation, all represent linear partial
differential equations with constant coefficients in $t$ and $\bm{r}$.

To find the electric 
field amplitude $\bm{ E}(\bm{ r},t)$ at time $t>0$,
we employ a spatial Fourier transform 
$\bm{ r} \rightarrow \bm{ k}$
and a temporal Laplace transform $t\rightarrow s$. This results
in a set of algebraic equations that connect the Laplace-Fourier transform
$\bm{ E}(\bm{k},s)$ of the electric field with the transforms of all
other dynamical fields.
The solution takes the form
\begin{align} 
  \bm{ E}(\bm{ k},s) &= \bm{ E}_0(\bm{ k})\cdot 
  \stackrel{\leftrightarrow}{R}(\bm{ k},s)\; ,
\label{eq:symbolicSol}\end{align} 
with matrix $ \stackrel{\leftrightarrow}{R}(\bm{ k},s)$ given in
Eq.~(\ref{eq:EperpRelat}).

Solution (\ref{eq:symbolicSol}) contains the complete information
about the propagation of radiation through an atom-plasma mixture.
Before studying the actual propagation in
Sec.~\ref{sec:pulsePropagation} we analyze the 
refractive index of the medium. The latter is usually represented 
as a complex function of the radiation frequency and can be 
derived from the roots of the denominator of
Eq.~(\ref{eq:symbolicSol}); see Sec.~\ref{sec:dispersion}.
In order to accomplish this we make a variable substitution
$s=-i\omega$. On the real axis, the (generally complex)  variable $\omega$ can be
interpreted as radiation frequency. The denominator of 
$ \stackrel{\leftrightarrow}{R}$ is then a function of $k$ and
$\omega$. Solving for its roots $k(\omega)$ and employing the relation
$k(\omega) = n(\omega) \omega/c$ would enable us to derive the
refractive index $n(\omega)$ of the mixture. However, we prefer to
use an equivalent approach, where  $k = n \omega/c$ 
represents variable substitution, and then solve directly for the roots
$n(\omega)$. Applying this variable substitution to $
\stackrel{\leftrightarrow}{R}$ yields
\begin{align} 
  \stackrel{\leftrightarrow}{R}\big (n,\omega, \hat{\bm{k}}\big) &= -
    \stackrel{\leftrightarrow}{{\cal M}}_0 \cdot
   \Big (
   \omega^2( n_\text{{\tiny A}}^2 -n^2)  \bm{ e}_{{ 1}}^{\, *} \otimes \bm{
    e}_{{ 1}}  
  + \omega^2 (1-n^2 )  \bm{ e}_{{ 2}}^{\, *} \otimes \bm{ e}_{{ 2}}  + 
   \stackrel{\leftrightarrow}{{\cal M}}
  \Big )^{-1},
\label{eq:EperpRelat2}
\end{align}
which is one of our main results. 
$n_\text{{\tiny A}}$ denotes the refractive index of the atomic gas in
the absence of the plasma. For two-level systems, it is derived in
App.~\ref{app:pureAtoms} and given by
\begin{equation} 
  n_\text{{\tiny A}}^2\big ( \omega,\hat{\bm{k}} \big ) = 
  1- \frac{ \eta \big (\hat{\bm{k}}\big)
  }{\omega-\omega_0 +\eta \big (\hat{\bm{k}}\big) +i \frac{ \Gamma}{2}}.
\label{eq:nAtResult}\end{equation} 
The parameter
\begin{align} 
   \eta \big (\hat{\bm{k}}\big) &:= 
 \frac{ {\cal N}_\text{{\tiny A}}}{\hbar\varepsilon_0} 
  \bm{d}_{\perp}\big (\hat{\bm{k}}\big)\cdot \bm{ d}_{eg}^{\, *}\; ,
\label{def:etak}\end{align}  
is related to the optical
cooperativity parameter \cite{PRA59:2427}. 
When $\eta \big (\hat{\bm{k}}\big) $ is significantly
larger than the decoherence rate,
the atomic density ${\cal N}_\text{{\tiny A}}$ is so large that the gas becomes opaque.
The two matrices $\stackrel{\leftrightarrow}{{\cal M}} $ and
$\stackrel{\leftrightarrow}{{\cal M}}_0 $ describe the influence of plasma electrons
on radiation propagation. Their form for a general classical plasma is
given by Eqs.~(\ref{def:Mmat}) and (\ref{def:M0}), respectively.

In our numerical examples we consider the case that all electrons 
are initially co-moving with velocity $\bm{ v}_\text{e} = \bm{ \beta} c$. 
If we separate the velocity vector into a 
longitudinal component  $\beta_\| = \hat{\bm{k}} \cdot \bm{ \beta} $
and a transverse part $\bm{ \beta}_{\perp} =
 \bm{\beta}- \hat{\bm{k}} ( \bm{\beta}\cdot \hat{\bm{k}})$, we 
can express the two matrices $\stackrel{\leftrightarrow}{{\cal M}} $ and
$\stackrel{\leftrightarrow}{{\cal M}}_0 $ as
\begin{align} 
   \stackrel{\leftrightarrow}{{\cal M}} &= 
  \frac{ \omega_\text{P}^2}{\gamma} 
  \left (
   \mathds{1} +
  \frac{\left (n^2-1+\frac{ \omega _\text{P}^2}{\gamma\omega^2}
   \right )  \bm{\beta}_\perp\otimes  \bm{\beta}_\perp
    }{ \left( 1-\beta _\| n\right)^2+\left(\beta
        _\|^2-1\right) 
  \frac{ \omega _\text{P}^2}{\gamma\omega^2} }
  \right ) ,
\label{eq:Msharp}\\  
    \stackrel{\leftrightarrow}{{\cal M}}_0  &=
     i \omega 
  \left ( n+1  - \frac{ \omega_\text{P}^2 }{\gamma\omega^2}
   \frac{\beta_\|  }{
    \left( 1-\beta _\| n\right)
   }
  \right ) \mathds{1} 
  + \frac{i \omega_\text{P}^2\left ( n
   \left( 1-\beta _\| n\right)
-\frac{ \omega_\text{P}^2}{\gamma\omega^2} \beta _\|  \right )
  \bm{\beta}_\perp\otimes  \bm{\beta}_\perp
   }{
   \gamma\omega\left( 1-\beta _\| n\right)
  \left( \left( 1-\beta _\| n\right)^2
  +\left(\beta _\|^2-1\right)
   \frac{ \omega_\text{P}^2}{\gamma\omega^2}\right)},
\label{eq:M0sharp}\end{align} 
with  $\omega_\text{P}^2=  q^2 {\cal N}_\text{e} /(m \varepsilon_0)$
the plasma frequency
and ${\cal N}_\text{e}$  the number density of plasma electrons.

Result (\ref{eq:EperpRelat2}) quantifies the physical effects that
we have discussed in the previous section. To explain this we first
remark that objects of the form $ \bm{v}^*\otimes \bm{v}$ for some
vector $\bm{v}$ are (proportional to) projectors that map any vector
to its component parallel to $\bm{v}$. Hence,
in absence of plasma 
($ \stackrel{\leftrightarrow}{{\cal M}} = 0$), the roots of
Eq.~(\ref{eq:EperpRelat2}) are given by $n=n_\text{A}$ ($n=1$) for radiation
with polarization $\bm{e}_{{1}}$ ($\bm{e}_{{2}}$),
respectively, so that only light with a
polarization along the atomic dipole moment will interact with the atoms.
The plasma electrons
interact strongly with radiation if they have a velocity $\beta_\| \approx
1/n$ because some terms in the denominator of Eqs.~(\ref{eq:Msharp})
and (\ref{eq:M0sharp}) then become small. If the electrons are not co-propagating with
the radiation pulse, $\bm{ \beta}_\perp \neq 0$, the terms proportional
to $\bm{\beta}_\perp\otimes  \bm{\beta}_\perp$ in Eqs.~(\ref{eq:Msharp})
and (\ref{eq:M0sharp}) alter the roots of Eq.~(\ref{eq:EperpRelat2}) and
describe the generation of volume plasmons.

Equation (\ref{eq:EperpRelat2}) enables us to draw
two further conclusions. First, the dependence of $ \stackrel{\leftrightarrow}{R}$
on the electron density always appears in form of a factor
$\omega_\text{P}^2/(\gamma \omega^2)$. The strength of the
plasma-radiation interaction is therefore determined by the ratio of
plasma frequency and optical frequency. In this paper, we
concentrate on underdense plasmas, for which $\omega_\text{P}<\omega$.
Second, we see that the direction of $\bm{\beta}_\perp$ determines
how atoms and plasma interact. If  $\bm{\beta}_\perp$ is parallel
to one of the polarization vectors $ \bm{ e}_{{ i}}$, then 
$ \stackrel{\leftrightarrow}{R}$ is a diagonal matrix. Two radiation
pulses with different polarization $ \bm{ e}_{{ 1}}$,
$ \bm{ e}_{{ 2}}$ then propagate independently.
On the other hand, if $\bm{\beta}_\perp$ is not parallel
to either polarization vector, the plasma induces a coupling
between both pulses.

\section{Optical dispersion relations and field amplitudes} \label{sec:dispersion}
For Laplace-Fourier transform (\ref{eq:symbolicSol}), the
spatiotemporal evolution of the radiation field is given by
\begin{align} 
  \bm{ E}(\bm{ r}, t) &= \int \frac{ d^3k}{ (2\pi)^{\frac{ 3}{2}}}\, e^{i \bm{ k}\cdot
    \bm{ r}}
  \int_{\cal P} \frac{ ds\, e^{ts} }{2\pi i}
\bm{ E}_0(\bm{ k})\cdot 
  \stackrel{\leftrightarrow}{R}(\bm{ k},s),
\label{eq:invLTFT}\end{align} 
where the path ${\cal P}$ is given by $s=r-i\omega$ for $\omega\in
(-\infty , \infty)$. The parameter $r$ has to be chosen so that the
path is to the right of all poles and branch cuts of the integrand.
Before we discuss the full evolution of a light pulse in
Sec.~\ref{sec:pulsePropagation}, it is worthwhile to consider a
plane-wave solution with fixed wavevector $\bm{k}$ and assume 
that for $t>0$ we can close the path ${\cal P}$ in the left half-plane
of $s$. The residue theorem then enables us to express the field as
\begin{align} 
  \bm{ E}(\bm{ r}, t) &\approx
  \sum_i   
 e^{i \bm{ k}\cdot \bm{ r}}  
    e^{t s_i(\bm{ k}) }  
  \bm{ E}_0(\bm{k})\cdot 
\stackrel{\leftrightarrow}{{\cal A}}(s_i(\bm{ k})).
\label{eq:wavepaketSuperposition}
\end{align} 
Here the sum runs over all poles $s_i(\bm{k})$, 
$ i=1,2,\ldots$ of the
function $\stackrel{\leftrightarrow}{R}(\bm{ k},s)$  in the 
left complex half-plane. By setting 
$s_i(\bm{ k})=-i\omega_i (\bm{k})$ we can study the dispersion
relation $\omega_i(\bm{k})$ associated with each pole.
The initial pulse is thus split into pulses with different dispersion
relations, which generally travel at different group velocities and
have different amplitudes  
$\stackrel{\leftrightarrow}{{\cal A}}(s_i)$, which are given by the residue of
$\stackrel{\leftrightarrow}{R}(\bm{ k},s)$ at pole $s_i$.
With this result we can obtain the refractive index of an atom-plasma mixture
by solving the equation $ \omega = \omega_i(\bm{k})$ for $k$ and then
setting $n(\omega)= c k(\omega)/\omega$.

In the non-relativistic limit and for electrons co-moving with the
laser beam, one simply adds the 
refractive indices of atomic gas and plasma. For electrons
moving in the plane spanned by $\hat{\bm{k}}$ and $\bm{e}_{{2}}$,
the optical properties of the medium are the same as for a pure plasma.
The most interesting case is 
when both $\bm{ \beta}_\perp$ and the radiation
polarization have a component along the direction 
$\bm{e}_{{ 1}}$ of the atomic dipole moment, so that volume
plasmons can be generated.
The field evolution is then
\begin{align} 
      \bm{ E}(\bm{ k},s=-i\omega) &= 
    - \frac{   {\cal M}_0 }{
  \omega^2( n_\text{{\tiny A}}^2 -n^2)  
  +   {\cal M}
  }
   \bm{ E}_0(\bm{ k}),
\label{eq:propIntCase}\end{align} 
where the complex numbers ${\cal M}$ and ${\cal M}_0$ take the form of 
Eqs.~(\ref{eq:Msharp}) and (\ref{eq:M0sharp}), respectively,
with $\mathds{1}$ replaced by 1 and 
$ \bm{\beta}_\perp\otimes  \bm{\beta}_\perp$ replaced
by $\beta_\perp^2$. 
The dispersion relations for radiation correspond to the poles
of the denominator of Eq.~(\ref{eq:propIntCase}), but the
analytical expressions are unwieldy and not presented here.
\begin{figure}
\begin{center}
(a) \includegraphics[width=7.5cm]{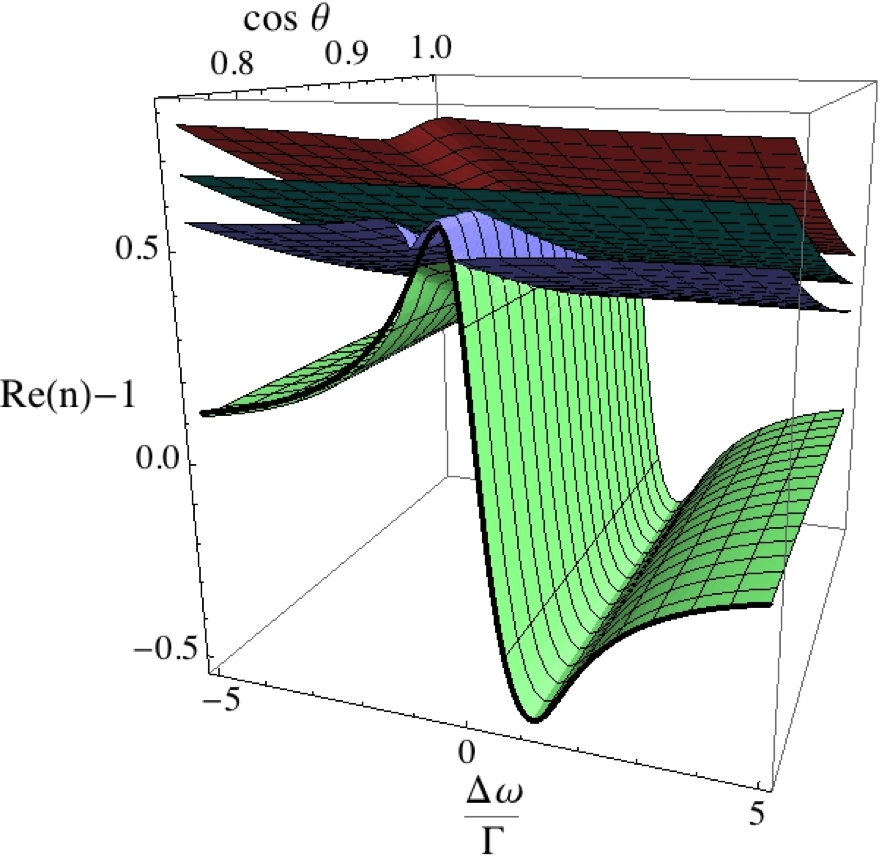}
\quad
(b) \includegraphics[width=7.5cm]{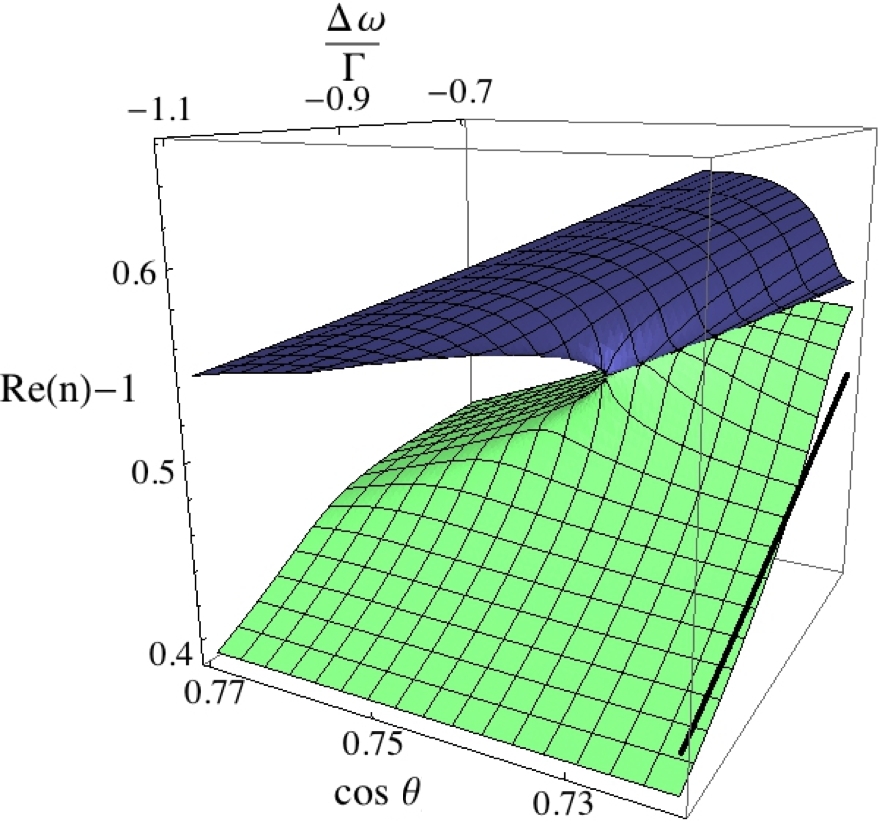}
\caption{\label{fig:dispRel}
Four dispersion relations, corresponding to the real part of the
refractive index, of radiation propagating through an atom-plasma
mixture. $\text{Re}(n)-1$ is displayed as a function of the detuning
$\Delta\omega =
\omega-\omega_0$ of the radiation beam from atomic resonance, and
of the angle $\theta$ between the radiation beam and the electron
velocity.
The black solid line displays the dispersion relation
in absence of a plasma. Figure (b) shows a detail of figure (a).} 
\end{center}
\end{figure}

Instead we discuss a specific numerical example. Experiments
with cold beams of ammonia molecules can typically achieve
 densities of ${\cal N}_\text{{\tiny A}} = 1.0\times 10^{9} \text{cm}^{-3}$ \cite{1367-2630-11-5-055030}.
Ammonia possesses an allowed electric-dipole transition at a resonance frequency
of 23.7 GHz, with a transition dipole matrix element of 
$ d_{eg} \approx 4.90\times 10^{-30}\text{Cm}$ \cite{10.1063/1.1617272}.
We assume that the experimental environment induces an
atomic decoherence rate of $\Gamma= 20\, \text{s}^{-1}$ and
work with a plasma electron density of ${\cal N}_\text{e} = 1.5\times 10^{11} \text{cm}^{-3}$.

For the numerical values above, Fig.~\ref{fig:dispRel}(a) shows 
the real part of the four poles, corresponding to four different
dispersion relations. Far away from the region where the four poles
are close to each other, they can be identified with specific physical phenomena.
The green refractive index is then close to that of
a pure atomic gas
(solid black line in Fig.~\ref{fig:dispRel}). The three poles that
form parallel sheets correspond to an electrostatic
wave that co-moves with the electrons at $n=\beta_\|^{-1}$, as well as
two poles with refractive index  
$n=\beta_\|^{-1} ( 1\pm \frac{ \omega_\text{P}}{\omega}
\gamma^{-\frac{ 3}{2}} )$, which are associated with the formation of volume plasmons.

\begin{figure}
\begin{center}
\includegraphics[width=7.0cm]{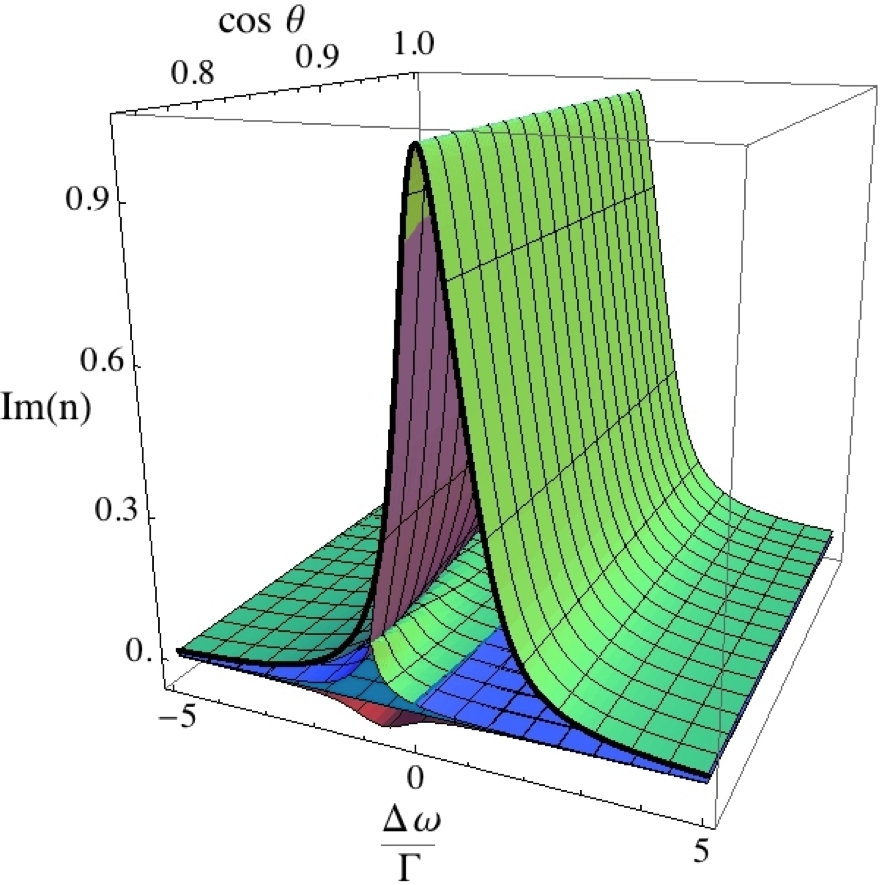}
\caption{\label{fig:absorb}
Absorption $\text{Im}(n)$ associated with the four dispersion
relations. The solid black line represents absorption in a pure
atomic gas.} 
\end{center}
\end{figure}

The effect of the plasma becomes particularly pronounced in the area 
where all poles are close, which happens when the resonance condition
$\beta_\| \approx n_\text{{\tiny A}}^{-1}$ is fulfilled. 
A detailed plot of this area is shown in  in Fig.~\ref{fig:dispRel}
b), which shows that the dispersion relations are strongly perturbed.
In particular, the red dispersion relation in Fig.~\ref{fig:absorb}
shows a gain (negative imaginary part) for small negative
detunings. This phenomenon has some relation to the
formation of wave instabilities in Rydberg gases \cite{Shukla2010},
but in the current case it is resonantly enhanced.
The energy for this process is provided by the kinetic energy of
plasma electrons. 

\begin{figure}
\begin{center}
\includegraphics[width=7.5cm]{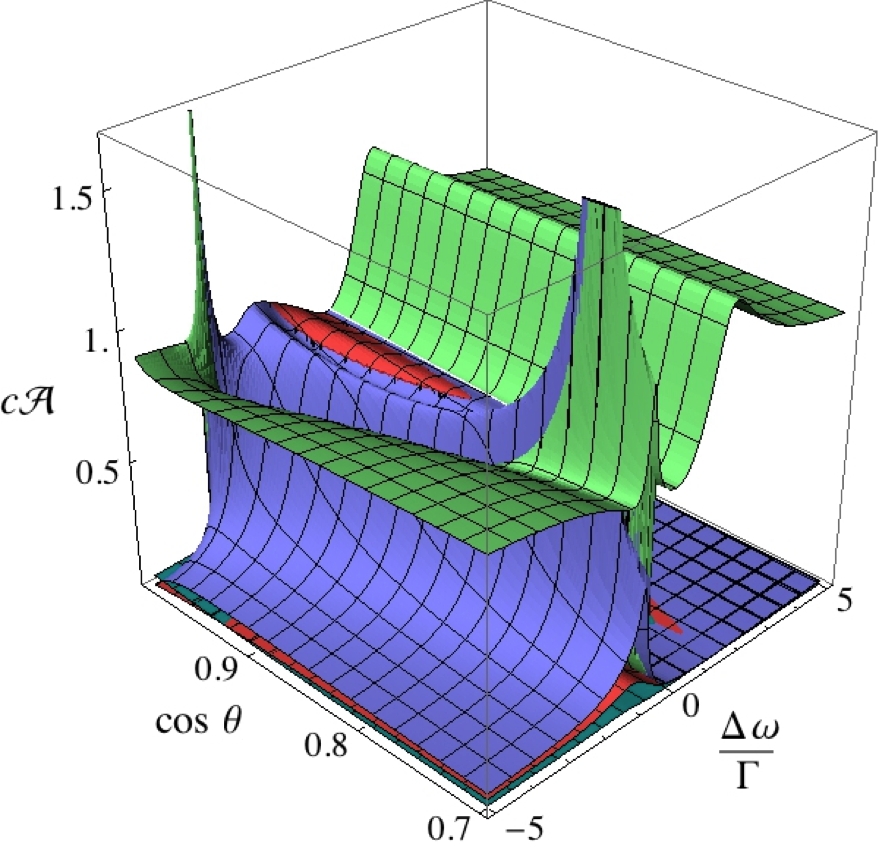}
\caption{\label{fig:amplitude}
Field amplitude factors ${\cal A}(s_i)$ of
Eq.~(\ref{eq:wavepaketSuperposition}), 
in units of $c$, for the four dispersion
relations. Sharp resonances occur at the points where two dispersion
relations meet.} 
\end{center}
\end{figure}

Of equal interest is the behaviour of the amplitude factors 
${\cal  A}(s_i)$, 
which multiply the partial
radiation pulses in Eq.~(\ref{eq:wavepaketSuperposition}). At points,
where the lower two dispersion relations of Fig.~\ref{fig:dispRel}
meet, the amplitude of the partial pulse is enhanced. In particular,
Fig.~\ref{fig:amplitude} shows that there are two points at which
amplitude resonances occur. At these points, both real and
imaginary part of the two dispersion relations are equal. 

To understand the appearance of these resonances, we first consider
an ideal, non-absorbing atom-plasma mixture with $\text{Im}(n_\text{{\tiny A}}) =0$.
For electrons moving at the speed $\beta_\|=n_\text{{\tiny A}}^{-1}$
the electrostatic dispersion relation $n=\beta_\|^{-1}$ matches the
atomic dispersion relation $n=n_\text{{\tiny A}}$, so that
both systems can be resonantly coupled.
However, for real atoms, $n_\text{{\tiny A}}$ always possesses
a non-vanishing imaginary part, see Eq.~(\ref{eq:nAtResult}),
so that a perfect match with the real dispersion relation
$n=\beta_\|^{-1}$ would appear impossible.
However, this argument neglects the influence
of volume plasmons, which can modify the imaginary part of both
dispersion relations. The strength of this effect depends on 
the ratio $r_p := \omega_\text{P}/(\omega \sqrt{\gamma})$, see the discussion
at the end of Sec.~\ref{sec:theoResults}.
It is therefore possible for two complex dispersion relations to take
the same values, if $r_p$ and $n_\text{{\tiny A}}$ (and thus $\Delta\omega$)
take specific values.

We have numerically evaluated under which conditions the two
resonances do appear. For the numerical parameters given above,
they only exist for $r_p$ between 0.05 and 0.2.
Hence, the phenomenon occurs only within a narrow range of
plasma electron densities. For $r_p \approx 0.05$,
both resonances are close together and appear at high electron
velocities $\beta_\| \approx 0.99$. As $r_p$ approaches the value
0.2, one resonance occurs at lower velocities, $\beta_\| \approx
0.75$, while the second resonance remains close to  
$\beta_\| \approx 0.99$. This is the case displayed in
Fig.~\ref{fig:amplitude}, where $r_p =0.12$.

\section{Pulse propagation}\label{sec:pulsePropagation}
In Sec.~\ref{sec:dispersion} we have identified
specific values of radiation frequency $\omega$ and
direction $\theta$ of the electrons for which the 
atom-plasma mixture acts like a resonant gain medium.
In this section we simulate the propagation of a radiation
pulse through such a medium.

We consider the situation that the pulse initially travels through a
vacuum, which fills a region in space characterized by $z<0$, and enters the
atom-plasma mixture (located in the region $z>0$) at a right angle.
The pulse is much wider than its wavelength, so that its transverse
profile does not change and the pulse is a function of $z$ and $t$
only. The polarization of the pulse is parallel to
the atomic dipole moment $\bm{d}_{eg}$ so that the interaction
with the atoms is maximized.

For mathematical reasons that are explained in
App.~\ref{app:pulseProp} we assume that, while inside the
vacuum, the pulse takes the form
\begin{align} 
  E_\text{free}(t,z) &= E_\text{peak}e^{-i\omega_c (t-t_0) + i  k_c z} \Pi
   \left(\frac{t-t_0-\frac{z}{c}}{T}\right) 
  \frac{1}{2} \left[1 + \cos \left(\frac{2 \pi}{T}
    \left(t-t_0-\frac{z}{c}\right)\right)\right],
\label{eq:initPulse}\end{align} 
with $\Pi(x)$ is the boxcar function, which is 1 for $|x|<0.5$ and
zero elsewhere.
$E_\text{free}(t,z)$ describes a pulse that oscillates
with central frequency $\omega_c$ and whose envelope is given by one 
period of the cosine function, so that it is similar to a Gaussian
function but nonzero only in a time interval of width $T$. 
 $k_c=\omega_c/c$ is the modulus of the central
wavevector $\bm{k}_c $ that is shown in Fig.~\ref{fig:sketch}.
The time $t_0$ can be chosen so that at $t=0$ the pulse is
completely inside the vacuum.

The pulse enters the atom-plasma mixture at $z=0$, where the boundary
conditions ensure continuity of $E(z,t)$. To find the evolution inside
the mixture we need to integrate Eq.~(\ref{eq:invLTFT}) numerically. 
We have seen in
Sec.~\ref{sec:dispersion} that if the electron velocity lies either
in the plane spanned by $\hat{\bm{k}} $ and $\bm{e}_{{1}}$ or
in the plane spanned by $\hat{\bm{k}} $ and $\bm{e}_{{2}}$,
then the electric field can be reduced to a scalar equation, as in
Eq.~(\ref{eq:propIntCase}) for instance.
We therefore only need to evaluate the 1D scalar form of
Eq.~(\ref{eq:invLTFT}) given by
\begin{align} 
  E(z, t) &=
 \int_{r-i\infty}^{r+i\infty} \frac{ ds\, e^{ts} }{2\pi i}
 \int_{-\infty}^\infty \frac{ dk}{ (2\pi)^{\frac{ 1}{2}}}\, e^{i k z}
  E_0(k)\, R(k,s).
\label{eq:invLTFT2}\end{align} 
For the parameters used in Sec.~\ref{sec:dispersion}, the numerical
integration of Eq.~(\ref{eq:invLTFT2})
is not feasible because the integrand varies significantly
on time scales $\Gamma^{-1}$, $\omega^{-1}$ and $\omega_P^{-1}$
that are many orders of magnitude apart.
To avoid this problem, and for improved presentation, we have doubled the electron density,
increased $\Gamma$ by a factor of $4\times 10^8$, and 
increased the atomic density by a factor of $3\times 10^8$, so that
$\Gamma$ and $\omega_P$ are in the order of
$0.1\, \omega$.  We will comment on the differences for pulses under
the more realistic conditions described in Sec.~\ref{sec:dispersion} below.
\begin{figure}[h]
\begin{center}
(a) \includegraphics[width=7cm]{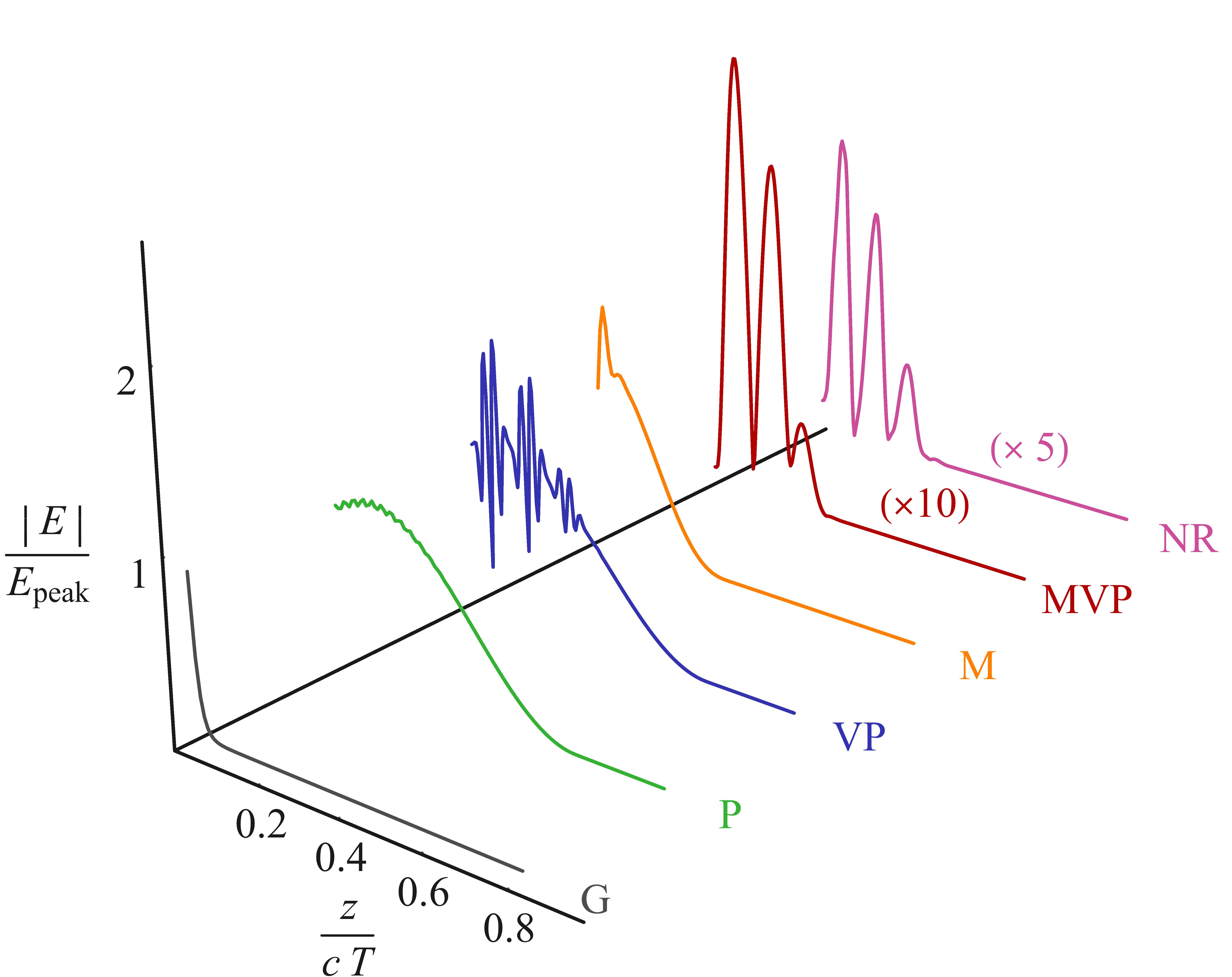}
\hspace{1cm}
(b) \includegraphics[width=7cm]{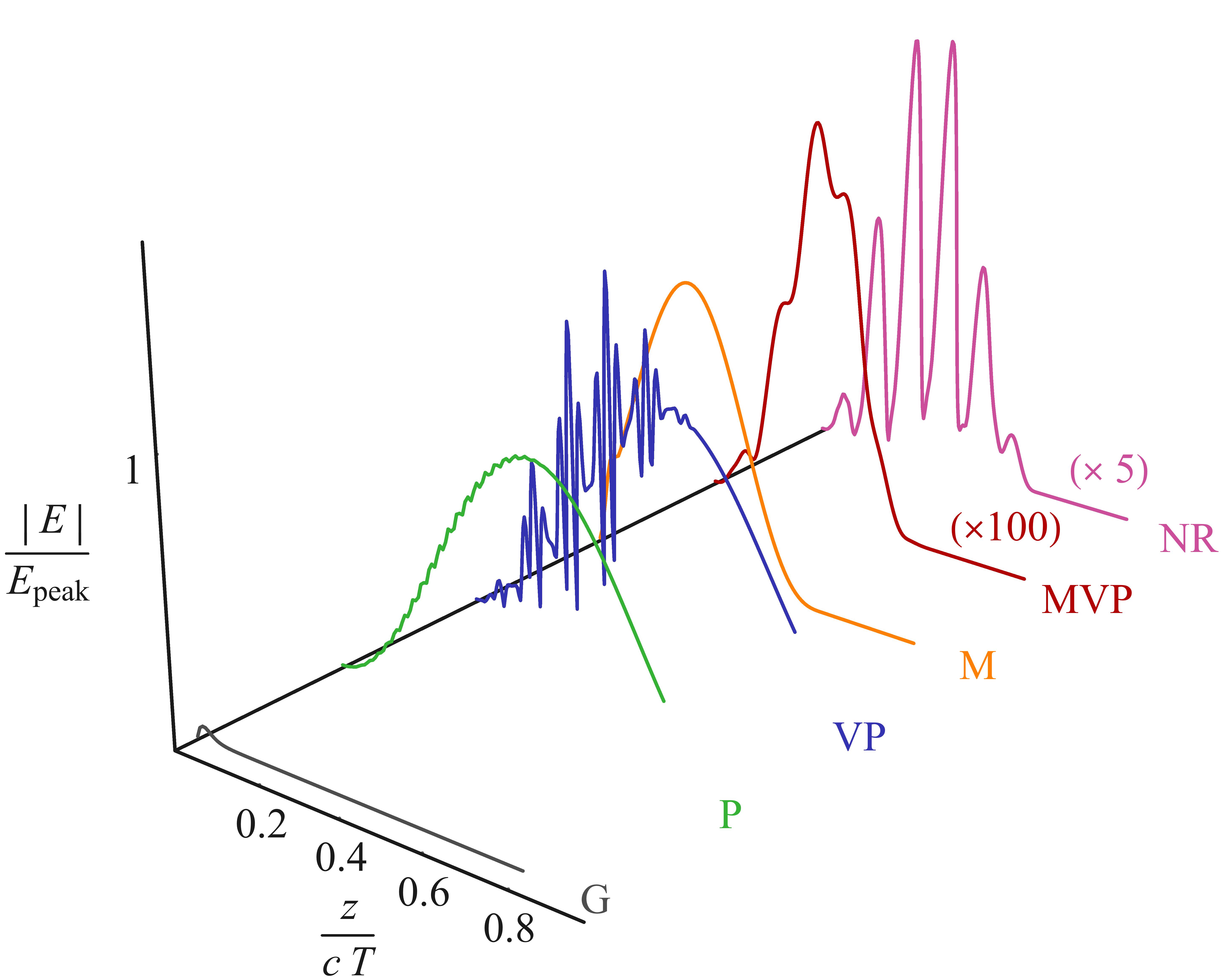}
\\[2mm]
 (c) \includegraphics[width=7cm]{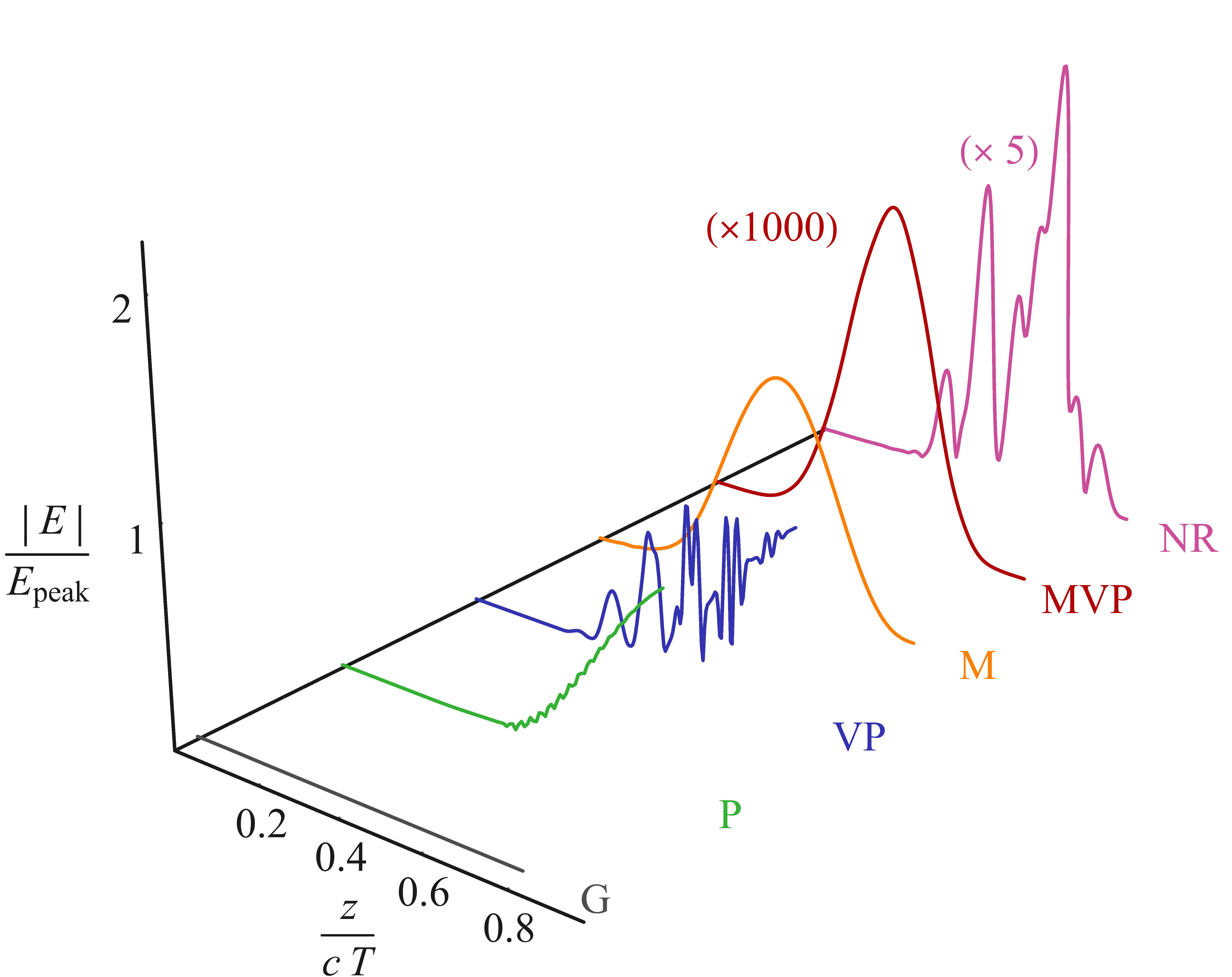}
\caption{\label{fig:pulseProp}
Propagation of light pulses through an atomic gas (G), 
a plasma (P), 
a plasma with volume plasmon generation (VP), 
an atom-plasma mixture (M),
an atom-plasma mixture with volume plasmon generation (MVP), 
and a non-resonant atom-plasma mixture (NR).
Shown is the electric field amplitude $|E(z,t)|$
in units of the peak amplitude $E_\text{peak}$ of the incoming pulse,
as a function of position $z$ in units of the pulse width $cT$. 
Figures a-c correspond to three different instances in time.
} 
\end{center}
\end{figure}

Our numerical results for this choice of parameters are shown in Fig.~\ref{fig:pulseProp}, which displays the pulse
for various media at three instances in time. 
Fig.~\ref{fig:pulseProp}(a) shows the pulse shortly after its peak entered the
medium, and Figs.~\ref{fig:pulseProp}(b) and (c) show the same pulse
at a time $0.4\, T$ and $0.8\, T$ later, where $T=20\, \Gamma^{-1}$ is the
duration of the pulse. 

The curve labelled ``G'' shows the pulse in an atomic gas in absence
of a plasma. The gas is strongly absorbing, so that the pulse
hardly enters the medium and is non-zero only for 
about $z <0.05\, cT$. In
Fig.~\ref{fig:pulseProp}(c) the pulse is completely absorbed.

The curve labelled ``P'' shows the pulse inside a pure plasma where
the electron velocity has no component along
$\bm{e}_{{1}}$, so that no plasmons are generated. The pulse has essentially the same
shape as in free space because $\omega_\text{P}$ is much smaller than $\omega$.
The small ripples that can be seen on the tail
of the pulse are an interference effect with a very small part of the
pulse that corresponds to an electrostatic wave.

The curve labelled ``VP'' corresponds to a pure plasma with volume plasmon generation.
As discussed in Sec.~\ref{sec:dispersion}, the pulse is then
decomposed into several components. The main part of
the pulse propagates in a similar way as in absence of plasmons, but
there are smaller pulse components that travel at different group
velocities and produce an interference pattern. 
In Fig.~\ref{fig:pulseProp}(c) one can see that at this instance in
time the slowest pulse components do not overlap with the main pulse 
anymore.

The curve labelled ``M'' displays an atom-plasma mixture without
plasmon generation. The main part of the pulse travels at a
group velocity that is comparable to that of the slow pulses in
the VP case. The narrow peak in Fig.~\ref{fig:pulseProp}(a)
corresponds to a pulse component that travels similarly to a pulse in
an atomic gas and is quickly absorbed. The peak of the main pulse is
increased, but only by about 10\%.

The curve labelled ``MVP'' is the main result of this section and displays a
pulse propagating through a mixture in the presence of plasmon
generation. We have numerically determined that a resonance occurs for an
electron direction of $\cos\theta \approx 0.714$ and a 
central pulse frequency of $\omega_c \approx 0.917\,\omega_0$.
These are the values used for all five curves discussed so far.
In Fig.~\ref{fig:pulseProp}(a) one can see that the pulse is split
into different components and strongly enhanced. For presentational
purposes we have rescaled the pulse:
 ``$(\times 10)$'' indicates that the field amplitude is actually 10 times
larger than displayed. Fig.~\ref{fig:pulseProp}(b) demonstrates that
the atom-plasma mixture acts as a gain medium for most of the pulse
components. At the instant in time in Fig.~\ref{fig:pulseProp}(c), the
pulse component with the largest gain factor dominates and is 
enhanced by a factor of about 2000. 
This demonstrates that 
atom-plasma mixtures may be able to amplify radiation. 

To clearly distinguish the effect of the resonance we have also
simulated a pulse labelled as ``NR'' that propagates in the same mixture but with
a different central frequency $\omega_c = 1.083\, \omega_0$,
which corresponds to the same detuning from $\omega_0$
as in case MVP, but with opposite sign. 
Initially, the pulse has a
similar shape as for the MVP case. Although it is smaller by a factor of
about two, it still is enhanced compared to the free pulse by a factor
of five. Fig.~\ref{fig:pulseProp}(b) shows that there is no further 
amplification of the pulse at this instance in time. In
Fig.~\ref{fig:pulseProp}(c) one can see that one pulse component is
enhanced. We attribute this late enhancement to the
fact that the spectrum of the pulse is not confined to a specific
frequency interval. Therefore, a small part of the pulse can still
fulfill the resonance criteria and is enhanced in the same way as
the red curve. However, the overall enhancement of the pulse 
lags behind the MVP case by a factor of about 200.

For the more realistic physical parameters used in
Sec.~\ref{sec:dispersion}, the pulse shape must be adapted to exhibit
amplification. The most relevant change is the pulse duration, which
must be in the order of $\Gamma^{-1}$ for resonant amplification and
is therefore much longer than for the results presented in this
section. However, the gain mechanism should work in a comparable
way because the effect of the plasma on the refractive index
is determined through the parameter $r_p = \omega_\text{P}/(\omega
\sqrt{\gamma})$. This parameter is similar for the parameters used
in Secs.~\ref{sec:dispersion} and \ref{sec:pulsePropagation}.

One can estimate the gain for the parameters used in
Sec.~\ref{sec:dispersion} through the imaginary part of the refractive
index displayed in Fig.~\ref{fig:absorb}. The negative value of
Im$(n) \approx -0.05$ for the red dispersion relation
generates an exponential gain factor 
$\exp(- \text{Im}(n) \omega z/c)$. This suggests that the radiation
field would grow by a factor of 2 over a propagation length
of 2.2 wavelengths, or 28 mm.

The results of this section demonstrate that volume plasmon generation
is indeed the mechanism behind radiation amplification in atom-plasma
mixtures. To achieve a high gain factor, the radiation pulse needs to
have a frequency close to the volume plasmon resonance and a
polarization that enables it to interact with the atoms.  The plasma
electrons must have an appropriate density, so that the plasma
frequency is about 10-20\% of the radiation frequency, 
and their velocity must be in a specific direction. If one of these
conditions is not met, the mixture will not amplify the radiation pulse.

\section{Conclusions}
We have studied the properties of a near-resonant radiation pulse propagating
through a mixture of two-level atoms and a 
classical, underdense, collision-less plasma.
If the plasma electrons have a velocity component in the direction
of the atomic dipole moment $\bm{d}_{eg}$, volume plasmons can be
formed and optical dispersion relations
are strongly modified, as shown in Fig.~\ref{fig:dispRel}.
For specific light frequencies $\omega$ and velocities of the electrons,
resonances occur (see Fig.~\ref{fig:pulseProp}) and the mixture acts as a gain medium
for radiation. This can happen if 
the plasma electrons travel at a speed comparable to the
velocity of light in the atomic gas, and if
the plasma frequency $\omega_\text{P}$ takes values between 
$0.05\, \omega$ and $0.2\, \omega$. We conclude from this
analysis that radiation amplification through generation 
of volume plasmons should be possible if the assumptions described in 
section \ref{sec:physDescr} are fulfilled.

However, more work is needed to understand whether 
light amplification via volume plasmons
could also be used for non-classical states of light without
destroying their coherence. Generating
volume plasmons itself is a coherent process, but there
are important secondary effects that have to be taken into account.

One such effect is the velocity distribution of plasma electrons,
which may lead to
inhomogenous broadening. For classical radiation, this
effect is included in the general results derived above. However, a
characterization of decoherence of a non-classical light pulse would
require an extension of our methods.
While decoherence due to inhomogeneous broadening could in
principle be reversed \cite{PhysRevLett.87.173601}, the corresponding 
procedure would likely reduce the gain factor.

Furthermore, collisions of plasma electrons with other particles and 
ionization of atoms are irreversible processes that
will decrease the coherence of a quantum state.
These processes are not included in this study and may
also modify the amplification of classical radiation pulses.
Further studies are necessary to provide a quantitative 
estimate of this effect. In future work we will
also investigate how the required speed of plasma electrons may
be reduced through a superposition of light pulses.

\acknowledgments

We gratefully acknowledge funding from ACEnet, AITF and NSERC.
KPM thanks St.~Francis Xavier University for a UCR grant.

\appendix
\section{Equations of motion}
\label{sec:dynnamicalEquations}
\subsection{Vlasov equation for relativistic plasmas}\label{sec:vlasov}
To describe plasma dynamics, we employ kinetic theory 
of a moving electron gas, in which the phase space density of
electrons $f(\bm{ r},\bm{ p},t) $ evolves according to the
relativistic Vlasov equation \cite{Liboff.R.L2003}
\begin{equation} 
  \partial_t f + \frac{ 1}{m\gamma} \bm{ p} \cdot \nabla_{\bm{r}} f + q
  \left [\bm{ E} + \frac{ 1}{m\gamma} \bm{ p}\times \bm{ B} \right]
   \cdot \nabla_{\bm{p}} f = \left . \frac{ df}{dt} \right |_\text{cl.}\!.
\end{equation} 
The right-hand side (r.h.s.) represents the contribution of
collisions between the particles, which we ignore in this paper.
Furthermore, we assume that the ions are not moving and that their spatial
distribution ${\cal N}_\text{I}$ is homogeneous.  Charge and current density 
of the plasma are then given by
$  \rho (\bm{ r},t) = q {\cal N}_\text{I} - q \int  d^3p\; f (\bm{ r},\bm{p},t)$
and $ \bm{ J}(\bm{ r},t) = - q \int d^3p\; \frac{ \bm{p}}{m\gamma}\, f
  (\bm{ r},\bm{ p},t)$, respectively.

For weak electromagnetic fields, the Vlasov equation can be linearized
in the deviation (\ref{eq:eqLinDist}) of the electron
distribution from the spatially homogeneous initial distribution $ f_0 (\bm{ p})$.
For quasi-neutral plasmas, where $ \int d^3p\; f_0 (\bm{ p}) = {\cal N}_\text{I}$, we obtain
Eq.~(\ref{eq:vlasov}) as dynamical equation.
Charge and current density take the form
\begin{align} 
  \rho (\bm{ r},t) &= - q \int d^3p\; \delta f (\bm{ r},\bm{ p},t),
\label{eq:chargeDensity}\\
  \bm{ J}(\bm{ r},t) &= - q \int d^3p\;  \frac{ 1}{m\gamma}\bm{ p}\, \delta f (\bm{
    r},\bm{ p},t)\; .
\label{eq:currentDensity}\end{align}  

Our main results are derived for general initial distributions 
$ f_0(\bm{ p}) $, but our numerical examples consider
plasma electrons with sharp momentum
$\bm{ p}_\text{e}=m \gamma  \bm{ v}_\text{e}$, so that 
\begin{equation} 
  f_0(\bm{ p}) = {\cal N}_\text{e}\delta(\bm{ p}-
    \bm{ p}_\text{e}) \; ,
\end{equation} 
with ${\cal N}_\text{e}$ the spatial electron density.

\subsection{Equations of motion for the atomic gas}
\label{sec:atomicDynamics} 
To describe the atomic gas we employ the method of coherence
operators $\hat{\sigma}_{ij}(\bm{ r})$ \cite{PRA65:022314}, which
are a set of operators that quantify superpositions
between two atomic states $|i \rangle , |j \rangle $. If their mean
value is zero, the atoms are in a state which does not include a
superposition of these states. For $i=j$, coherence operators
$\hat{\sigma}_{ii}(\bm{ r})$ describe the population of the atomic state
$|i \rangle $. 
Formally, coherence operators can be defined as
\begin{align} 
  \hat{\sigma}_{ij}(\bm{ r}) &:= \frac{ 1}{{\cal N}_\text{A}} \int d^3 r'\,
  \hat{\Psi}_i(\bm{ r}')^\dagger\,   
  \hat{\Psi}_j(\bm{ r}') \, S(\bm{ r}- \bm{ r}') \, , 
\label{eq:cohFieldOp}\end{align} 
where $\hat{\Psi}_i(\bm{ r})$ is an atomic field operator that
annihilates an atom in internal state $|i \rangle $ at position
$\bm{r}$ \cite{PhysRevA.49.3799,PhysRevA.50.1681}.
$S(\bm{ r}- \bm{ r}')$ is a smooth
non-negative function that is zero if $\bm{r}' \notin V_S(\bm{r})$,
where $V_S(\bm{r})$ is an area of volume $V_S$ around $\bm{r}$. The
function $S$ is approximately given by 
$S(\bm{ r}- \bm{ r}') \approx V_S^{-1}$ for $\bm{r}' \in V_S(\bm{r})$
and drops rapidly to zero around the boundary of $V_S(\bm{r})$, such that
$ \int d^3 r\, S(\bm{r}) = 1$. In microscopic quantum electrodynamics
one usually sets 
$S(\bm{ r}- \bm{ r}') = \delta(\bm{ r}- \bm{r}')$. 
However, because we are employing macroscopic electrodynamics that
is averaged over length scales large compared to atoms but small
compared to the wavelength \cite{JacksonEdyn}, the same averaging
has to be applied to all dynamical fields. This is accomplished
through $S(\bm{ r}- \bm{ r}')$ in Eq.~(\ref{eq:cohFieldOp}).
For a homogeneous atomic gas that initially is prepared in ground
state $|g \rangle $, the atomic density is given by
$ {\cal N}_\text{A} = \langle  \hat{\Psi}_g(\bm{r})^\dagger\,
  \hat{\Psi}_g(\bm{r}) \rangle $.

The dynamics of coherence operators described by
the Heisenberg equation of motion. In the presence
of incoherent processes such as spontaneous emission, the
Heisenberg-Langevin equation is used. Both require to evaluate the
commutator, which is given by
\begin{align} 
  [\hat{\sigma}_{ij}(\bm{r})  ,  \hat{\sigma}_{kl}(\bm{r}') ] &=
  \frac{\delta(\bm{r}- \bm{r}')}{{\cal N}_\text{A}}  
  \left \{
  \delta_{il} \sigma_{kj}(\bm{r})
  -  \delta_{kj} \sigma_{il}(\bm{r})
  \right \} .
\end{align} 
Strictly speaking, the $\delta$-distribution in this expression should
be replaced by $S(\bm{r}- \bm{r}')$. However, on the length scales of
macroscopic electrodynamics, $S(\bm{r}- \bm{r}')$ can be considered 
as a representation of the  $\delta$-distribution.

Within the dipole approximation for atoms, coherence
  operators can provide a complete dynamical description, but because we are
  interested in the semi-classical properties 
of radiation propagation we restrict our considerations to the mean
value $\sigma_{ij}(\bm{ r})$  of the
coherence fields and ignore field fluctuations.
For two-level atoms, the equation relevant for the optical
properties of the gas is then given by 
\begin{align} 
  \partial_t \sigma_{eg}(\bm{ r},t) &=
  \left  (-i \omega_0 -\frac{ \Gamma}{2}\right )\sigma_{eg}(\bm{
    r},t) 
 +\frac{i}{\hbar\varepsilon_0}   
  \bm{D}_\perp(\bm{ r}) \cdot  \bm{ d}_{eg}^{\, *} (\sigma_{gg}(\bm{ r})
  - \sigma_{ee}(\bm{ r}) )\; ,
\end{align} 
with $\Gamma$ the atomic spectral line width,
$ \bm{ d}_{eg}$ the electric-dipole moment of the atoms, and $\bm{D}$
the electric displacement field. 
The notation $(\cdots)_\perp$ indicates that only the transverse part of
a vector field is considered. In Fourier space, the transverse part of
an arbitrary vector field $\bm{ V}(\bm{ k})$ can be found using
\begin{equation} 
  V_{\perp, i} (\bm{ k}) = \left (\delta_{ij} - \hat{\bm{k}}_i \hat{k_j}
  \right ) V_j(\bm{ k})\; ,
\label{eq:vectorTransversalization}\end{equation} 
with $ \hat{\bm{k}} $ of Eq.~(\ref{def:e1}).

For small radiation intensities we can make the
low-saturation approximation by setting
$\sigma_{gg}(\bm{ r}) \approx 1$ and $\sigma_{ee}(\bm{ r}) \approx 0$.
In rotating-wave approximation \cite{AtomPhotonInteractions},
the atomic dynamics is then described by
Eq.~(\ref{eq:atomdynamics}).

\subsection{Coupled equations of motion}
The atomic dynamical equation (\ref{eq:atomdynamics}) and the
Vlasov equation (\ref{eq:vlasov}) are coupled through an
electromagnetic field, which evolves according to the
macroscopic Maxwell equations. In the associated material
equation $ \bm{ D} = \varepsilon_0 \bm{ E} + \bm{ P}$, the
polarization field contains the contribution of bound charges, i.e.,
the atoms and molecules, 
and is given by the transverse part of Eq.~(\ref{eq:atomPolariz}). 
The longitudinal part can
be neglected because it is only off-resonantly coupled to the
radiation.
A spatial Fourier transform $\bm{ r} \rightarrow \bm{ k}$,
and a temporal Laplace transform $t\rightarrow s$  
of all dynamical equations yields
\begin{align} 
     \delta f (\bm{ k},\bm{ p}, s)  &= \frac{ 1}{s+ i \frac{ \bm{ k} \cdot \bm{
         p} }{m\gamma} } \bigg [ \delta f(\bm{ k},\bm{ p}, t=0) + q
    \Big (  \bm{E} (\bm{ k},s)
   +  \frac{ 1}{m\gamma}\bm{
        p}\times \bm{ B}  (\bm{ k},s) \Big )  \cdot \nabla_{\bm{p}} f_0 (\bm{ p})
  \bigg ],
\label{eq:fp}\\
 \sigma_{eg}(\bm{k},s) &= \frac{ 
    \sigma_{eg}(\bm{k},t=0) +\frac{i}{\hbar\varepsilon_0}   
  \bm{D}_\perp(\bm{k},s) \cdot  \bm{ d}_{eg}^{\, *}  }{s+ i \omega_0 +\frac{ \Gamma}{2}} ,
\label{eq:sigma}\\
  i \bm{ k} \cdot \bm{ D}(\bm{k},s) &= \rho (\bm{k},s),
\\
  i \bm{ k}\cdot \bm{ B}(\bm{k},s) &= 0,
\\
  i \bm{ k} \times \bm{ E}(\bm{k},s) &= - s\bm{ B}(\bm{k},s) +\bm{ B}(\bm{k},t=0),
\\
  i \bm{ k} \times \bm{ B}(\bm{k},s) &= \mu_0 
 \left (\bm{ J}(\bm{k},s) + s \bm{ D}(\bm{k},s)-\bm{
     D}(\bm{k},t=0)
 \right).
\end{align} 
In rotating-wave approximation, the material equations take the
explicit form
\begin{align} 
    \bm{ D}_{\perp} (\bm{k},t) &= \varepsilon_0
    \bm{ E}_{\perp}(\bm{k},t) + 
  {\cal N}_\text{{\tiny A}} \, \bm{d}_{\perp} (\hat{\bm{k}})
  \sigma_{eg}(\bm{ k})  ,
\label{eq:material}\\
  D_{\|} (\bm{k},t) &= \varepsilon_0
    E_{\|i}(\bm{k},t),
\end{align} 
with $  \bm{d}_{\perp} (\hat{\bm{k}})$ of Eq.~(\ref{eq:dPerpDef}).

\section{Solving the equations of motion}\label{app:solution}
We assume that atoms are initially in their ground state, 
$\sigma_{eg} (t=0) =0$, and that the initial state of the plasma
electrons is given by $f_0$, so that 
$  \delta f(\bm{ k},\bm{ p}, t=0)=0$. 
The incoming electromagnetic
field (radiation pulse) is characterized by the initial amplitudes
$\bm{ B}_0(\bm{ k}) :=\bm{ B}(\bm{ k},t=0) $ and
$\bm{D}_0(\bm{ k}) :=\bm{D}(\bm{ k},t=0) $. 

Using Eq.~(\ref{eq:material}), the 
electromagnetic field can be
expressed in terms of $\sigma_{eg}$ and $\delta f$ as
\begin{align} 
  \bm{ B}(\bm{ k},s) &= \frac{ 1}{s} \left (
   \bm{ B}_0(\bm{ k}) -i \bm{ k}\times   \bm{ E}_\perp(\bm{ k},s) 
  \right ) ,
\label{eq:B1}\\
  E_\parallel(\bm{ k},s)  &= -\frac{ i}{k\varepsilon_0} \rho(\bm{
    k},s) ,
\\
  \bm{ E}_\perp (\bm{ k},s)  &= \frac{ 1}{c^2 k^2 +  s^2}
  \Big (
    \frac{ s}{\varepsilon_0} \left ( \bm{D}_0(\bm{k})
  -  \bm{J}_{\perp}(\bm{k},s) \right )
  +  i c^2 \bm{k}\times  \bm{B}_0(\bm{k})
  -\frac{ s^2 {\cal N}_\text{{\tiny A}}}{\varepsilon_0} 
  \bm{d}_{\perp} \big (\hat{\bm{k}}\big)\, \sigma_{eg}(\bm{k},s)
  \Big  ),
\label{eq:Eperp1}\\
  \bm{ D}_\perp (\bm{ k},s)  &= \frac{ 1}{c^2 k^2 +  s^2}
  \Big (
    s \left ( \bm{D}_0(\bm{k})
  -  \bm{J}_{\perp}(\bm{k},s) \right ) 
  +\frac{  i}{\mu_0}  \bm{k}\times  \bm{B}_0(\bm{k})
  + k^2 c^2  {\cal N}_\text{{\tiny A}} 
   \bm{d}_{\perp} \big (\hat{\bm{k}}\big)\, \sigma_{eg}(\bm{k},s)
  \Big ) \; .
\label{eq:Dperp1}\end{align}
Inserting Eq.~(\ref{eq:Dperp1}) into
Eq.~(\ref{eq:sigma}) we obtain
\begin{align} 
  \sigma_{eg}(\bm{ k},s) &= \frac{ 
\left (
   i s \left (  \bm{D}_0(\bm{k}) - \bm{J}_{\perp}(\bm{k},s)
   \right )
   -\frac{ \bm{k}\times  \bm{B}_0(\bm{k})}{\mu_0} 
  \right ) \cdot \bm{ d}_{eg}^{\, *} }{\hbar \varepsilon_0\Delta(\bm{ k},s)} ,
\label{eq:sigmaDeltaRWA} 
\\
  \Delta(\bm{ k},s) &:= (c^2 k^2 +  s^2) \left (s+ i \omega_0
    +\frac{ \Gamma}{2} \right )
  -i k^2 c^2 \eta \big (\hat{\bm{k}}\big) ,
\label{def:Delta}
\end{align} 
with $\eta \big (\hat{\bm{k}}\big) $ of Eq.~(\ref {def:etak}).
Inserting the atomic polarization (\ref{eq:sigmaDeltaRWA})  into
Eq.~(\ref{eq:Eperp1}) yields
\begin{align} 
  \bm{ E}_\perp (\bm{ k},s)  &=  \frac{ 1}{\varepsilon_0} \frac{
  \left ( \bm{ X}(\bm{ k},s)- s \bm{J}_{\perp}(\bm{k},s) 
  \right )\cdot  \stackrel{\leftrightarrow}{{\cal
      E}^{(\text{A})}}(\bm{ k},s)
  }{c^2 k^2 +  s^2}  ,
\label{eq:Eperp2}\\
  \bm{ X}(\bm{ k},s) &:= \frac{ i}{\mu_0} \bm{ k}\times\bm{B}_0(\bm{k}) + s  \bm{D}_0(\bm{k}),
\\
  {\cal E}^{(\text{A})}_{ij} (\bm{ k},s) &:= \delta_{ij} -i\frac{  s^2}{\Delta(\bm{ k},s)} \frac{
    {\cal N}_\text{{\tiny A}}}{\hbar\varepsilon_0 } \bm{ d}_{eg,i}^{\, *} \bm{
    d}_{\perp ,j}\big (\hat{\bm{k}}\big)\; .
\label{def:EA}\end{align} 
The tensor $ {\cal E}^{(\text{A})} $ describes the
polarization-dependent 
response of the atomic medium to the radiation field. 
$\bm{ X}$ depends only on the initial radiation field.
Inserting Eq.~(\ref{eq:Eperp2}) into (\ref{eq:fp}) yields
\begin{align} 
       \delta f (\bm{ k},\bm{ p}, s)  &= \frac{ q}{\varepsilon_0} \Bigg \{
   \frac{  \left ( \bm{ X}(\bm{ k},s)- s \bm{J}_{\perp}(\bm{k},s) 
  \right )\cdot  \stackrel{\leftrightarrow}{{\cal
      E}^{(\text{A})}}(\bm{ k},s)  }{c^2 k^2 + s^2}
 + \frac{ \bm{ p}\times
  \left ( \bm{ k}\times  
   \left ( \bm{ X}- s \bm{J}_{\perp} 
  \right )\cdot  \stackrel{\leftrightarrow}{{\cal
      E}^{(\text{A})}}(\bm{ k},s)
  \right )}{i s m \gamma (c^2 k^2 + s^2)}
\nonumber \\  &\hspace{4mm}
 +\frac{ \rho(\bm{ k},s) }{i k^2} \bm{ k} 
  +\varepsilon_0  \frac{ \bm{ p}\times \bm{B}_0(\bm{k})
  }{s m\gamma}
 \Bigg \} \cdot \frac{\nabla_{\bm{p}} f_0 (\bm{ p})}{s+i \frac{ \bm{ k}\cdot
     \bm{ p}}{m\gamma} }.
\label{eq:fp2}\end{align} 

Eq.~(\ref{eq:fp2}) could be used to derive charge and current density
distribution. However, it is physically more instructive to 
use Eq.~(\ref{eq:fp}) instead. Multiplying it with $-q$
and integrating over $d^3 p$ gives a relation between charge
distribution and radiation field,
\begin{align} 
  \rho(\bm{ k},s) &= -q^2 \bm{ E}(\bm{ k},s)\cdot \int d^3 p\, 
  \frac{  \nabla_{\bm{p}} f_0(\bm{ p})}{s+i \bm{ k}\cdot \bm{
      p}/(m\gamma)}
   -q^2 \bm{ B}(\bm{ k},s)\cdot \int 
   \frac{ d^3 p}{m\gamma s + i \bm{ k}\cdot \bm{p}} 
 \nabla_{\bm{p}} f_0(\bm{ p})\times \bm{ p}\; .
\label{eq:rhoInterm}\end{align} 
Exploiting $\nabla_{\bm{p}} m\gamma = \bm{ p} /(m\gamma c^2)$ and
performing a partial integration gives
\begin{align} 
  \rho(\bm{ k},s) &=  \varepsilon_0 \bm{ E}(\bm{ k},s)\cdot 
  \left (
  i \bm{ k} \partial_s  {\cal Z}(\bm{ k},s) - \frac{ 1}{c^2} 
   (\bm{ k}\cdot \nabla_{\bm{k}})\bm{ {\cal Z}}(\bm{ k},s)
  \right ) 
   + \varepsilon_0 \bm{ B}(\bm{ k},s)\cdot 
  \left ( \bm{ k}\times  \nabla_{\bm{k}}  {\cal Z} (\bm{ k},s) \right )\; ,
\label{eq:rho1}\end{align} 
where we have introduced the notation
\begin{align} 
  {\cal Z}(\bm{ k},s) &:=\frac{ q^2 }{\varepsilon_0}\int \frac{ d^3 p}{m \gamma}\;
  \frac{ f_0(\bm{ p})}{s+i \bm{ k}\cdot \bm{ p}/(m \gamma)},
\\
   \bm{ {\cal Z}}(\bm{ k},s) &:=\frac{ q^2 }{\varepsilon_0} \int \frac{ d^3 p}{(m \gamma)^2}\;
  \frac{ f_0(\bm{ p})}{s+i \bm{ k}\cdot \bm{ p}/(m \gamma)} \bm{
    p},
\\
     \Omega_\text{P}^2&:=\frac{ q^2 }{\varepsilon_0} \int
    \frac{ d^3 p}{m \gamma}\; f_0(\bm{ p}) \; .
\end{align}
$  {\cal Z}$ is a relativistic generalization of the plasma dispersion
function, which takes the same form with $\gamma=1$
\cite{SitenkoMalnev1995}.  
$\bm{ {\cal Z}}$ is related to
the dispersion function through
$ i \partial_s  \bm{ {\cal Z}} =  \nabla_{\bm{k}} {\cal Z}$.
The relativistic plasma frequency
$ \Omega_\text{P}$, which coincides with the standard
plasma frequency $\omega_\text{P}$ in the limit $\gamma\rightarrow 1$,
fulfills
$  i \bm{ k}\cdot \bm{ {\cal Z}}  = \Omega_\text{P}^2 - s  {\cal Z} $.

Relation $\bm{ E}(\bm{ k},s)\cdot \bm{ k}  = k E_\parallel = -i \rho
/\varepsilon_0$ enables us to express the charge density as
\begin{align} 
 \rho(\bm{ k},s)= \frac{\zeta \varepsilon_0 }{ 1 -
   \partial_s {\cal Z} } 
  \Bigg (
  \frac{ 1}{c^2} \bm{ E}(\bm{ k},s)\cdot \left (
   \bm{ {\cal Z}} + s \partial_s  \bm{ {\cal Z}} \right )
  -i \left (\bm{ k}\times  \bm{ B}(\bm{ k},s)\right )
  \cdot \partial_s \bm{  {\cal Z}} 
  \Bigg )
   \; .
\label{eq:rho2} \end{align} 
In this result, we introduced a parameter $\zeta$, which enables us 
to evaluate the non-relativistic limit.
$\zeta = 1$ corresponds to the full relativistic treatment, while
$\zeta=0$ neglects the coupling to the magnetic field and 
relativistic corrections.

The current density (\ref{eq:currentDensity}) can be found
in a similar way. Using Eq.~(\ref{eq:fp}) again we find
\begin{align} 
   \bm{ J}(\bm{ k},s) &=  \varepsilon_0 \bm{ E}(\bm{ k},s)   {\cal Z}
   + \rho(\bm{ k},s)
   \partial_s  \bm{ {\cal Z}}- 
  \zeta \varepsilon_0 \bm{ B}(\bm{ k},s)
   \times \bm{ {\cal Z}}
  + i \zeta\varepsilon_0 \left [ \left (
  i \bm{ k}\times  \bm{ B}(\bm{k},s) 
  -\frac{ s}{c^2}  \bm{ E}(\bm{k},s)
\right )\cdot \nabla_{\bm{k}}\right ]  \bm{{\cal  Z}}.
\label{eq:J1} \end{align} 

Solving the system of equations 
(\ref{eq:rho2}), (\ref{eq:J1}), (\ref{eq:Eperp2}), and
(\ref{eq:B1}) is a straightforward but tedious task.
The first step is to express charge and current density
in terms of transverse electromagnetic fields, 
\begin{align} 
  \rho(\bm{ k},s) &=\zeta \frac{ \varepsilon_0}{s}\left (
    \rho_0 (\bm{ k},s) + 
   \bm{  \rho}_1(\bm{k},s) \cdot  \bm{E}_\perp (\bm{ k},s)  \right ),
\\
    \bm{ J}_\perp(\bm{ k},s) &= \frac{  \varepsilon_0}{s} \left ( \zeta \bm{ J}_0 (\bm{ k},s) +
  \bm{E}_\perp (\bm{ k},s)  \cdot
   \stackrel{\leftrightarrow}{{\cal M}} \right ),
\\
  \rho_0 (\bm{ k},s)  &:=\frac{ - i}{ 
   ({\cal Z}_\perp +1) } (\bm{ k}\times
  \bm{ B}_0) \cdot \partial_s \bm{ {\cal Z}},
\\
   \bm{  \rho}_1(\bm{k},s)   &:=
  \frac{1 }{c^2({\cal Z}_\perp+1) } \bm{ U},
\\
   (\bm{ J}_0 (\bm{ k},s))_n  &:=
  -  (\bm{ k}\times \bm{ B}_0)_i 
  \Big [ i \frac{ \bm{ U}_n \,\partial_s \bm{{\cal Z}}_{\perp ,i}}{
      c^2 k^2({\cal Z}_\perp+1)}
  + \left (  \frac{ \partial \bm{ {\cal Z}}_n}{\partial k_i} 
  \right  )_\perp 
  \Big ] 
  + \frac{ i (s {\cal Z}-\Omega_\text{p}^2)}{k^2 }  (\bm{
    k}\times \bm{ B}_0)_n ,
\end{align} 
with
\begin{align} 
  {\cal M}_{in} &:= \zeta \frac{ U_i U_n}{
     k^2 c^4({\cal Z}_\perp+1) }
  -i \zeta \frac{ k^2 c^2 + s^2}{c^2} 
  \left (  \frac{ \partial \bm{ {\cal Z}}_n}{\partial k_i} 
  \right  )_\perp 
  +
  \left (\zeta \Omega_\text{P}^2 +(1-\zeta)s {\cal Z} \right)
  \left ( \delta_{in}  \right  )_\perp ,
\label{def:Mmat}\\
  \bm{ U} &:= (k^2 c^2 + s^2) \partial_s \bm{ {\cal Z}}_\perp + s
  \bm{ {\cal Z}}_\perp ,
\\
  {\cal Z}_\perp &:= 
  \frac{ 1}{k^2 c^2} \left (
  \Omega_\text{P}^2 -(k^2 c^2 + s^2) \partial_s {\cal Z} - 2 s {\cal Z}
  \right ).
\end{align} 
These quantities describe the dependence of the optical properties
on the plasma electron density, temperature, and mean velocity.
The notation $(\cdots )_\perp$ implies again that all indices have been
transversalized according to Eq.~(\ref{eq:vectorTransversalization}).
The final result for the electric field then takes the form
(\ref{eq:symbolicSol}), with 
\begin{align} 
   \stackrel{\leftrightarrow}{R}(\bm{ k},s) 
   &=  -
    \stackrel{\leftrightarrow}{{\cal M}}_0 \cdot
   \left (
  (k^2 c^2+s^2)  \stackrel{\leftrightarrow}{{\cal E}^{(\text{A})}}^{-1}  + 
   \stackrel{\leftrightarrow}{{\cal M}}
  \right )^{-1} ,
\label{eq:EperpRelat}\\
 \left (   \stackrel{\leftrightarrow}{{\cal M}}_0
  \right )_{in} &:=\frac{ i\zeta }{ k c^3 (1+{\cal
      Z}_\perp)} \partial_s \bm{ {\cal Z}}_{\perp,i} U_n + 
   \zeta \frac{k}{c}  \left (  
  \frac{ \partial \bm{ {\cal Z}}_n}{\partial k_i} 
  \right  )_\perp 
+ \frac{ i}{ck} 
  \left ( ck (ck+is)-\zeta s {\cal Z}+\zeta \Omega_\text{P}^2 \right )
  \left ( \delta_{in}  \right  )_\perp.
\label{def:M0}\end{align} 
In this expression, we have assumed that the initial field amplitudes
$\bm{ B}_0(\bm{ k})$ and $\bm{E}_0(\bm{ k}) $
are transverse vector fields, are
are related as in free space through 
$ \bm{k}\times  \bm{B}_0(\bm{k})
  = - \bm{E}_0(\bm{k}) \, k/c$. 

\subsection{Pure atomic gas}\label{app:pureAtoms}
Expression (\ref{eq:EperpRelat}) can be cast into a 
physically more intuitive form if one relates 
$ \stackrel{\leftrightarrow}{{\cal E}^{(\text{A})}}$ to
the atomic refractive index $n_\text{{\tiny A}}$. To do this, we
temporarily assume that the plasma density vanishes, so that
\begin{align} 
   \bm{ E}(\bm{ k},s) &=  - i 
   \frac{  ck+is}{k^2 c^2+s^2} \bm{ E}_0(\bm{ k})\cdot 
  \stackrel{\leftrightarrow}{{\cal E}^{(\text{A})}}\; .
\label{eq:EperpNoPlasma}
\end{align} 
Here, and in the more general expression (\ref{eq:EperpRelat}), all
vectors that multiply 
$ \stackrel{\leftrightarrow}{{\cal E}^{(\text{A})}}$ from the left
are transverse. We therefore can replace 
$ \stackrel{\leftrightarrow}{{\cal E}^{(\text{A})}}$
by its transverse part $ {\cal E}^{(\text{A})\perp}$, which is
found using Eq.~(\ref{eq:vectorTransversalization}).
This matrix has three left-eigenvectors, which are given
by $\hat{\bm{k}} $, $\bm{ e}_{{ 1}}$ and
$\bm{ e}_{{ 2}}$ of Eq.~(\ref{def:e1}).
The eigenvalue of $\hat{\bm{k}} $ is zero because $ {\cal
  E}^{(\text{A})\perp}$ is a transverse tensor, and the eigenvalue of
$\bm{ e}_{{ 2}} \big (\hat{\bm{k}}\big)$ is unity, because 
the atoms are transparent
for radiation with
this polarization.
For radiation with polarization $\bm{ e}_{{ 1}} \big (\hat{\bm{k}}\big) $,
the interaction with the atoms is described through the relation
\begin{align} 
 \bm{ e}_{{ 1}} \cdot {\cal
   E}^{(\text{A})\perp} (\bm{ k},s) &= 
 \left (1 -i\frac{  s^2 \eta\big(\hat{\bm{k}}\big)}{\Delta(\bm{ k},s)} 
  \right )  \bm{ e}_{{ 1}}.
\label{eq:activeLightPolarization}
\end{align} 

For radiation with initial amplitude 
$\bm{ E}_0(\bm{ k})= E_0(\bm{ k})\, \bm{ e}_{{ 1}}$,  the
evolution equation becomes
\begin{align} 
   E(\bm{ k},s) &= 
  \frac{ (s-i c k) \left(\Gamma -2 i \eta +2 s+2 i \omega _0\right)}{
   2 i \omega _0\left (c^2 k^2+s^2\right)+c^2 k^2 (\Gamma -2 i \eta +2 s)+s^2
   (\Gamma +2 s)}
    E_0(\bm{ k}).
\label{eq:EAdef}\end{align} 
The roots of the denominator determine the dispersion relation. Setting
$s=-i\omega$ and $k = n_\text{{\tiny A}}(\omega) \omega/c$, and solving for
the atomic refractive index we find $n_\text{{\tiny A}}$ of Eq.~(\ref{eq:nAtResult}). 

Conversely, we can express $\eta \big(\hat{\bm{k}}\big)$ of Eq.~(\ref{def:etak})
through the refractive index,
\begin{align} 
  \eta\big(\hat{\bm{k}}\big) &=
  \left( 1 - \frac{ 1}{
   n_\text{{\tiny A}}^2\big(\omega,\hat{\bm{k}}\big)} \right) 
  \left(-i \frac{ \Gamma}{2}-(\omega -\omega _0)\right).
\end{align} 
If we reinsert this expression into Eq.~(\ref{def:EA}) we get 
\begin{align} 
  {\cal E}^{(\text{A})\perp} (\bm{ k},s) &=
       \bm{ e}_{{ 1}}^{\, *} \otimes \bm{ e}_{{ 1}} 
   \frac{ \omega^2-k^2 c^2}{ n_\text{{\tiny A}}^2(\omega,\hat{
       k})\,\omega^2-k^2 c^2}
   +
    \bm{ e}_{{ 2}}^{\, *} \otimes \bm{ e}_{{ 2}}.
\label{eq:EAperp2}\end{align} 

Returning to the propagator 
(\ref{eq:EperpRelat}) in the presence of a plasma,
the dynamics of the electric field amplitude takes the form
(\ref{eq:EperpRelat2})

\section{Propagation of radiation pulses}\label{app:pulseProp}
Our task is to evaluate the pulse amplitude for $t>0$
inside the atom-plasma mixture ($z>0$). Under these circumstances, the factor
$e^{i k z}$ in Eq.~(\ref{eq:invLTFT2}) guarantees that the integrand decreases to zero for 
$|k|\rightarrow \infty$ and Im$(k)>0$ as long as the factor
$ E_0(k)\, R(k,s)$ does not increase indefinitely.
In this case we can close the $k$-integral in the upper half-plane
and use the Residue theorem for its evaluation.

Due to the fact that $R(k,s)$ is a known rational function, the
requirement that $ E_0(k)\, R(k,s)$ remains finite poses a condition
on the shape of the initial pulse $ E_0(z)$, which for instance
excludes a Gaussian shape of the initial pulse. We have verified
that our method can indeed not be applied to
Gaussian pulses and therefore have chosen an initial pulse of the 
form (\ref{eq:initPulse}).

We can then use the Residue theorem to evaluate the
$k$-integration. denoting by $k_l(s), l=1,2,\cdots , L$ the poles of the
integrand in the upper half of the complex $k$-plane, the field can be
expressed as  
\begin{align} 
  E(z, t) &=
 \frac{1}{ \sqrt{2\pi}} \int_{r-i\infty}^{r+i\infty}  ds\, e^{ts} 
   \sum_l e^{i k_l(s) z} E_0(k_l(s))  R_l(s),
\label{eq:EfieldProp1}\end{align} 
where $  R_l(s) = \text{Res} \left ( R(k,s), k\rightarrow k_l(s)
\right )$ denotes the residue of $R(k,s)$ at the $l$th pole.

The integral over $s=r-i\omega$ can be evaluated numerically, but one has to
choose the parameter $r$ carefully. To understand this, consider a
simple example of a medium with some given refractive index $n^2(s) =
1 +\chi(s)$, where $\chi(s)$ is the Laplace transform of the linear
susceptibility $\chi(t)$. For such a medium, the denominator of $R(k,s)$
takes the form $k^2 c^2 + s^2 n^2(s)$, so that the poles are at
$k_l(s) =\pm i s n(s)/c$. If we consider the usual situation where
Re$(n)>0$, then only the pole $k_l(s) = i s n(s)/c$ propagates to the
right. Let us assume for now that $s=-i\omega$ for real frequencies
$\omega$.
For an absorbing medium we have Im$(n)>0$ so that the pole
lies in the upper half and one recovers the usual result for a pulse
propagating through an absorbing medium. However, for a gain medium
the pole would be in the lower half and therefore would not contribute
to Eq.~(\ref{eq:EfieldProp1}). We remark that this problem is not
resolved by the Kramers-Kronig relations. The latter imply that all poles
of $n(s)$ are in the lower half plane, but this does not restrict
the location of the poles $\pm i s n(s)/c$ in complex $k$-plane.

This problem can be resolved by recalling that $s=r-i\omega$ has a
positive real part $r$ that has to be larger than the real part of all
poles in the $s$-plane. However, increasing $r$ also affects
the pole $k_l(s) = i s n(s)/c$ in the $k$-plane and increases its
imaginary part. For large enough $r$ one can achieve that 
the imaginary parts of all poles $k_l(s) $ (for which the real part is
positive so that they propagate to the right) are positive and thus
contribute to Eq.~(\ref{eq:EfieldProp1}). In our numerical
evaluations,
we have used a value of $r=0.0173\, \omega_0$ for all pulses displayed in
Fig.~\ref{fig:pulseProp}.

The last step is to implement the proper initial conditions for $E(z,
t)$. For the situation considered in this section, the boundary
conditions at the interface between mixture and vacuum are that
$E(z,t)$ and its derivatives with respect to $z$ are continuous at $z=0$.
Eq.~(\ref{eq:EfieldProp1}) implies that the temporal Laplace transform of
$E(z,t)$ is given by 
\begin{align} 
  E(z, s) &=
\sqrt{2\pi} i  \sum_l e^{i k_l(s) z} E_0(k_l(s))  R_l(s) .
\label{eq:EfieldProp2}\end{align} 
We can then form a set of $L$ coupled linear equations
\begin{align} 
   \left . \partial_z^l E(z, s) \right |_{z=0} 
  &=
   \left . \partial_z^l E_\text{free}(z, s) \right |_{z=0} 
 \quad , \; l=1, 2, \cdots , L
\end{align} 
and solve this for the unknown parameters $ E_0(k_l(s))$, thus
expressing the relevant initial conditions $E_0(k_l(s))$ through the
incoming pulse $ E_\text{free}(z, t)$. It then only remains to
numerically integrate the resulting integral over $s$, which can be
done with standard algorithms provided by Mathematica.

\bibliography{UNP_article1.bib}

\begin{thebibliography}{49}%
\makeatletter
\providecommand \@ifxundefined [1]{%
 \@ifx{#1\undefined}
}%
\providecommand \@ifnum [1]{%
 \ifnum #1\expandafter \@firstoftwo
 \else \expandafter \@secondoftwo
 \fi
}%
\providecommand \@ifx [1]{%
 \ifx #1\expandafter \@firstoftwo
 \else \expandafter \@secondoftwo
 \fi
}%
\providecommand \natexlab [1]{#1}%
\providecommand \enquote  [1]{``#1''}%
\providecommand \bibnamefont  [1]{#1}%
\providecommand \bibfnamefont [1]{#1}%
\providecommand \citenamefont [1]{#1}%
\providecommand \href@noop [0]{\@secondoftwo}%
\providecommand \href [0]{\begingroup \@sanitize@url \@href}%
\providecommand \@href[1]{\@@startlink{#1}\@@href}%
\providecommand \@@href[1]{\endgroup#1\@@endlink}%
\providecommand \@sanitize@url [0]{\catcode `\\12\catcode `\$12\catcode
  `\&12\catcode `\#12\catcode `\^12\catcode `\_12\catcode `\%12\relax}%
\providecommand \@@startlink[1]{}%
\providecommand \@@endlink[0]{}%
\providecommand \url  [0]{\begingroup\@sanitize@url \@url }%
\providecommand \@url [1]{\endgroup\@href {#1}{\urlprefix }}%
\providecommand \urlprefix  [0]{URL }%
\providecommand \Eprint [0]{\href }%
\providecommand \doibase [0]{http://dx.doi.org/}%
\providecommand \selectlanguage [0]{\@gobble}%
\providecommand \bibinfo  [0]{\@secondoftwo}%
\providecommand \bibfield  [0]{\@secondoftwo}%
\providecommand \translation [1]{[#1]}%
\providecommand \BibitemOpen [0]{}%
\providecommand \bibitemStop [0]{}%
\providecommand \bibitemNoStop [0]{.\EOS\space}%
\providecommand \EOS [0]{\spacefactor3000\relax}%
\providecommand \BibitemShut  [1]{\csname bibitem#1\endcsname}%
\let\auto@bib@innerbib\@empty
\bibitem [{\citenamefont {Chang}\ \emph {et~al.}(2007)\citenamefont {Chang},
  \citenamefont {S{\o}rensen}, \citenamefont {Demler},\ and\ \citenamefont
  {Lukin}}]{lukin:NatPhys2007}%
  \BibitemOpen
  \bibfield  {author} {\bibinfo {author} {\bibfnamefont {D.~E.}\ \bibnamefont
  {Chang}}, \bibinfo {author} {\bibfnamefont {A.~S.}\ \bibnamefont
  {S{\o}rensen}}, \bibinfo {author} {\bibfnamefont {E.~A.}\ \bibnamefont
  {Demler}}, \ and\ \bibinfo {author} {\bibfnamefont {M.~D.}\ \bibnamefont
  {Lukin}},\ }\href {\doibase 10.1038/nphys708} {\bibfield  {journal} {\bibinfo
   {journal} {Nature Physics}\ }\textbf {\bibinfo {volume} {3}},\ \bibinfo
  {pages} {807 } (\bibinfo {year} {2007})}\BibitemShut {NoStop}%
\bibitem [{\citenamefont {Nie}\ and\ \citenamefont
  {Emory}(1997)}]{Science275-1102}%
  \BibitemOpen
  \bibfield  {author} {\bibinfo {author} {\bibfnamefont {S.}~\bibnamefont
  {Nie}}\ and\ \bibinfo {author} {\bibfnamefont {S.~R.}\ \bibnamefont
  {Emory}},\ }\href {\doibase 10.1126/science.275.5303.1102} {\bibfield
  {journal} {\bibinfo  {journal} {Science}\ }\textbf {\bibinfo {volume}
  {275}},\ \bibinfo {pages} {1102 } (\bibinfo {year} {1997})}\BibitemShut
  {NoStop}%
\bibitem [{\citenamefont {Schmidt}\ and\ \citenamefont {Imamo{\v
  g}lu}(1996)}]{Schmidt96}%
  \BibitemOpen
  \bibfield  {author} {\bibinfo {author} {\bibfnamefont {H.}~\bibnamefont
  {Schmidt}}\ and\ \bibinfo {author} {\bibfnamefont {A.}~\bibnamefont {Imamo{\v
  g}lu}},\ }\href {http://ol.osa.org/abstract.cfm?id=45256} {\bibfield
  {journal} {\bibinfo  {journal} {Opt. Lett.}\ }\textbf {\bibinfo {volume}
  {21}},\ \bibinfo {pages} {1936} (\bibinfo {year} {1996})}\BibitemShut
  {NoStop}%
\bibitem [{\citenamefont {Hau}\ \emph {et~al.}(1999)\citenamefont {Hau},
  \citenamefont {Harris}, \citenamefont {Dutton},\ and\ \citenamefont
  {Behroozi}}]{Hau99}%
  \BibitemOpen
  \bibfield  {author} {\bibinfo {author} {\bibfnamefont {L.~V.}\ \bibnamefont
  {Hau}}, \bibinfo {author} {\bibfnamefont {S.~E.}\ \bibnamefont {Harris}},
  \bibinfo {author} {\bibfnamefont {Z.}~\bibnamefont {Dutton}}, \ and\ \bibinfo
  {author} {\bibfnamefont {C.~H.}\ \bibnamefont {Behroozi}},\ }\href {\doibase
  10.1038/175611} {\bibfield  {journal} {\bibinfo  {journal} {Nature}\ }\textbf
  {\bibinfo {volume} {397}},\ \bibinfo {pages} {594} (\bibinfo {year}
  {1999})}\BibitemShut {NoStop}%
\bibitem [{\citenamefont {Kang}\ and\ \citenamefont {Zhu}(2003)}]{Kang03}%
  \BibitemOpen
  \bibfield  {author} {\bibinfo {author} {\bibfnamefont {H.}~\bibnamefont
  {Kang}}\ and\ \bibinfo {author} {\bibfnamefont {Y.}~\bibnamefont {Zhu}},\
  }\href {\doibase 10.1103/PhysRevLett.91.093601} {\bibfield  {journal}
  {\bibinfo  {journal} {Phys. Rev. Lett.}\ }\textbf {\bibinfo {volume} {91}},\
  \bibinfo {eid} {093601} (\bibinfo {year} {2003})}\BibitemShut {NoStop}%
\bibitem [{\citenamefont {Wang}\ \emph {et~al.}(2006)\citenamefont {Wang},
  \citenamefont {Marzlin},\ and\ \citenamefont {Sanders}}]{wang:063901}%
  \BibitemOpen
  \bibfield  {author} {\bibinfo {author} {\bibfnamefont {Z.-B.}\ \bibnamefont
  {Wang}}, \bibinfo {author} {\bibfnamefont {K.-P.}\ \bibnamefont {Marzlin}}, \
  and\ \bibinfo {author} {\bibfnamefont {B.~C.}\ \bibnamefont {Sanders}},\
  }\href {\doibase 10.1103/PhysRevLett.97.063901} {\bibfield  {journal}
  {\bibinfo  {journal} {Phys. Rev. Lett.}\ }\textbf {\bibinfo {volume} {97}},\
  \bibinfo {eid} {063901} (\bibinfo {year} {2006})}\BibitemShut {NoStop}%
\bibitem [{\citenamefont {Guerreiro}\ \emph {et~al.}(2014)\citenamefont
  {Guerreiro}, \citenamefont {Martin}, \citenamefont {Sanguinetti},
  \citenamefont {Pelc}, \citenamefont {Langrock}, \citenamefont {Fejer},
  \citenamefont {Gisin}, \citenamefont {Zbinden}, \citenamefont {Sangouard},\
  and\ \citenamefont {Thew}}]{PhysRevLett.113.173601}%
  \BibitemOpen
  \bibfield  {author} {\bibinfo {author} {\bibfnamefont {T.}~\bibnamefont
  {Guerreiro}}, \bibinfo {author} {\bibfnamefont {A.}~\bibnamefont {Martin}},
  \bibinfo {author} {\bibfnamefont {B.}~\bibnamefont {Sanguinetti}}, \bibinfo
  {author} {\bibfnamefont {J.~S.}\ \bibnamefont {Pelc}}, \bibinfo {author}
  {\bibfnamefont {C.}~\bibnamefont {Langrock}}, \bibinfo {author}
  {\bibfnamefont {M.~M.}\ \bibnamefont {Fejer}}, \bibinfo {author}
  {\bibfnamefont {N.}~\bibnamefont {Gisin}}, \bibinfo {author} {\bibfnamefont
  {H.}~\bibnamefont {Zbinden}}, \bibinfo {author} {\bibfnamefont
  {N.}~\bibnamefont {Sangouard}}, \ and\ \bibinfo {author} {\bibfnamefont
  {R.~T.}\ \bibnamefont {Thew}},\ }\href {\doibase
  10.1103/PhysRevLett.113.173601} {\bibfield  {journal} {\bibinfo  {journal}
  {Phys. Rev. Lett.}\ }\textbf {\bibinfo {volume} {113}},\ \bibinfo {pages}
  {173601} (\bibinfo {year} {2014})}\BibitemShut {NoStop}%
\bibitem [{\citenamefont {Kulin}\ \emph {et~al.}(2000)\citenamefont {Kulin},
  \citenamefont {Killian}, \citenamefont {Bergeson},\ and\ \citenamefont
  {Rolston}}]{Kulin2000}%
  \BibitemOpen
  \bibfield  {author} {\bibinfo {author} {\bibfnamefont {S.}~\bibnamefont
  {Kulin}}, \bibinfo {author} {\bibfnamefont {T.~C.}\ \bibnamefont {Killian}},
  \bibinfo {author} {\bibfnamefont {S.~D.}\ \bibnamefont {Bergeson}}, \ and\
  \bibinfo {author} {\bibfnamefont {S.~L.}\ \bibnamefont {Rolston}},\ }\href
  {\doibase 10.1103/PhysRevLett.85.318} {\bibfield  {journal} {\bibinfo
  {journal} {Phys. Rev. Lett.}\ }\textbf {\bibinfo {volume} {85}},\ \bibinfo
  {pages} {318} (\bibinfo {year} {2000})}\BibitemShut {NoStop}%
\bibitem [{\citenamefont {Fletcher}\ \emph {et~al.}(2006)\citenamefont
  {Fletcher}, \citenamefont {Zhang},\ and\ \citenamefont
  {Rolston}}]{Fletcher2006}%
  \BibitemOpen
  \bibfield  {author} {\bibinfo {author} {\bibfnamefont {R.~S.}\ \bibnamefont
  {Fletcher}}, \bibinfo {author} {\bibfnamefont {X.~L.}\ \bibnamefont {Zhang}},
  \ and\ \bibinfo {author} {\bibfnamefont {S.~L.}\ \bibnamefont {Rolston}},\
  }\href@noop {} {\bibfield  {journal} {\bibinfo  {journal} {Phys. Rev. Lett.}\
  }\textbf {\bibinfo {volume} {96}},\ \bibinfo {pages} {105003} (\bibinfo
  {year} {2006})}\BibitemShut {NoStop}%
\bibitem [{\citenamefont {Simien}\ \emph {et~al.}(2004)\citenamefont {Simien},
  \citenamefont {Chen}, \citenamefont {Gupta}, \citenamefont {Laha},
  \citenamefont {Martinez}, \citenamefont {Mickelson}, \citenamefont {Nagel},\
  and\ \citenamefont {Killian}}]{Simien2004}%
  \BibitemOpen
  \bibfield  {author} {\bibinfo {author} {\bibfnamefont {C.~E.}\ \bibnamefont
  {Simien}}, \bibinfo {author} {\bibfnamefont {Y.~C.}\ \bibnamefont {Chen}},
  \bibinfo {author} {\bibfnamefont {P.}~\bibnamefont {Gupta}}, \bibinfo
  {author} {\bibfnamefont {S.}~\bibnamefont {Laha}}, \bibinfo {author}
  {\bibfnamefont {Y.~N.}\ \bibnamefont {Martinez}}, \bibinfo {author}
  {\bibfnamefont {P.~G.}\ \bibnamefont {Mickelson}}, \bibinfo {author}
  {\bibfnamefont {S.~B.}\ \bibnamefont {Nagel}}, \ and\ \bibinfo {author}
  {\bibfnamefont {T.~C.}\ \bibnamefont {Killian}},\ }\href {\doibase
  10.1103/PhysRevLett.92.143001} {\bibfield  {journal} {\bibinfo  {journal}
  {Phys. Rev. Lett.}\ }\textbf {\bibinfo {volume} {92}},\ \bibinfo {pages}
  {143001} (\bibinfo {year} {2004})}\BibitemShut {NoStop}%
\bibitem [{\citenamefont {Killian}\ \emph {et~al.}(2005)\citenamefont
  {Killian}, \citenamefont {Chen}, \citenamefont {Gupta}, \citenamefont {Laha},
  \citenamefont {Martinez}, \citenamefont {Mickelson}, \citenamefont {Nagel},
  \citenamefont {Saenz},\ and\ \citenamefont {Simien}}]{Killian2005}%
  \BibitemOpen
  \bibfield  {author} {\bibinfo {author} {\bibfnamefont {T.~C.}\ \bibnamefont
  {Killian}}, \bibinfo {author} {\bibfnamefont {Y.~C.}\ \bibnamefont {Chen}},
  \bibinfo {author} {\bibfnamefont {P.}~\bibnamefont {Gupta}}, \bibinfo
  {author} {\bibfnamefont {S.}~\bibnamefont {Laha}}, \bibinfo {author}
  {\bibfnamefont {Y.~N.}\ \bibnamefont {Martinez}}, \bibinfo {author}
  {\bibfnamefont {P.~G.}\ \bibnamefont {Mickelson}}, \bibinfo {author}
  {\bibfnamefont {S.~B.}\ \bibnamefont {Nagel}}, \bibinfo {author}
  {\bibfnamefont {A.~D.}\ \bibnamefont {Saenz}}, \ and\ \bibinfo {author}
  {\bibfnamefont {C.~E.}\ \bibnamefont {Simien}},\ }\href {\doibase doi:
  10.1088/0953-4075/38/2/026} {\bibfield  {journal} {\bibinfo  {journal} {J.
  Phys. B}\ }\textbf {\bibinfo {volume} {38}},\ \bibinfo {pages} {S351}
  (\bibinfo {year} {2005})}\BibitemShut {NoStop}%
\bibitem [{\citenamefont {Mendon\c{c}a}\ \emph {et~al.}(2009)\citenamefont
  {Mendon\c{c}a}, \citenamefont {J.Loureiro},\ and\ \citenamefont
  {Ter\c{c}as}}]{Mendonca2009}%
  \BibitemOpen
  \bibfield  {author} {\bibinfo {author} {\bibfnamefont {J.~T.}\ \bibnamefont
  {Mendon\c{c}a}}, \bibinfo {author} {\bibnamefont {J.Loureiro}}, \ and\
  \bibinfo {author} {\bibfnamefont {H.}~\bibnamefont {Ter\c{c}as}},\ }\href
  {\doibase doi:10.1017/S0022377809007971} {\bibfield  {journal} {\bibinfo
  {journal} {J. Plasma Phys.}\ }\textbf {\bibinfo {volume} {75}},\ \bibinfo
  {pages} {713} (\bibinfo {year} {2009})}\BibitemShut {NoStop}%
\bibitem [{\citenamefont {Shukla}(2010)}]{Shukla2010}%
  \BibitemOpen
  \bibfield  {author} {\bibinfo {author} {\bibfnamefont {P.}~\bibnamefont
  {Shukla}},\ }\href {\doibase 10.1016/j.physleta.2010.07.013} {\bibfield
  {journal} {\bibinfo  {journal} {Phys. Lett. A}\ }\textbf {\bibinfo {volume}
  {374}},\ \bibinfo {pages} {3656–3657} (\bibinfo {year} {2010})}\BibitemShut
  {NoStop}%
\bibitem [{\citenamefont {Lu}\ \emph {et~al.}(2011)\citenamefont {Lu},
  \citenamefont {Guo}, \citenamefont {Guo},\ and\ \citenamefont
  {Han}}]{Ronghua2011}%
  \BibitemOpen
  \bibfield  {author} {\bibinfo {author} {\bibfnamefont {R.}~\bibnamefont
  {Lu}}, \bibinfo {author} {\bibfnamefont {L.}~\bibnamefont {Guo}}, \bibinfo
  {author} {\bibfnamefont {J.}~\bibnamefont {Guo}}, \ and\ \bibinfo {author}
  {\bibfnamefont {S.}~\bibnamefont {Han}},\ }\href {\doibase
  doi:10.1016/j.physleta.2011.04.027} {\bibfield  {journal} {\bibinfo
  {journal} {Phys. Lett. A}\ }\textbf {\bibinfo {volume} {375}},\ \bibinfo
  {pages} {2158ﾐ2161} (\bibinfo {year} {2011})}\BibitemShut {NoStop}%
\bibitem [{\citenamefont {Mendon\c{c}a}\ \emph {et~al.}(2010)\citenamefont
  {Mendon\c{c}a}, \citenamefont {Shukla},\ and\ \citenamefont
  {Shukla}}]{Mendonca2010}%
  \BibitemOpen
  \bibfield  {author} {\bibinfo {author} {\bibfnamefont {J.~T.}\ \bibnamefont
  {Mendon\c{c}a}}, \bibinfo {author} {\bibfnamefont {N.}~\bibnamefont
  {Shukla}}, \ and\ \bibinfo {author} {\bibfnamefont {P.~K.}\ \bibnamefont
  {Shukla}},\ }\href {\doibase doi:10.1017/S0022377809990389} {\bibfield
  {journal} {\bibinfo  {journal} {J. Plasma Phys.}\ }\textbf {\bibinfo {volume}
  {76}},\ \bibinfo {pages} {19} (\bibinfo {year} {2010})}\BibitemShut {NoStop}%
\bibitem [{\citenamefont {Baranger}\ and\ \citenamefont
  {Mozer}(1961)}]{Baranger1961}%
  \BibitemOpen
  \bibfield  {author} {\bibinfo {author} {\bibfnamefont {M.}~\bibnamefont
  {Baranger}}\ and\ \bibinfo {author} {\bibfnamefont {B.}~\bibnamefont
  {Mozer}},\ }\href {\doibase 10.1103/PhysRev.123.25} {\bibfield  {journal}
  {\bibinfo  {journal} {Phys. Rev.}\ }\textbf {\bibinfo {volume} {123}},\
  \bibinfo {pages} {25} (\bibinfo {year} {1961})}\BibitemShut {NoStop}%
\bibitem [{\citenamefont {Kunze}\ and\ \citenamefont
  {Griem}(1968)}]{Kunze1968}%
  \BibitemOpen
  \bibfield  {author} {\bibinfo {author} {\bibfnamefont {H.~J.}\ \bibnamefont
  {Kunze}}\ and\ \bibinfo {author} {\bibfnamefont {H.~R.}\ \bibnamefont
  {Griem}},\ }\href {\doibase 10.1103/PhysRevLett.21.1048} {\bibfield
  {journal} {\bibinfo  {journal} {Phys. Rev. Lett.}\ }\textbf {\bibinfo
  {volume} {21}},\ \bibinfo {pages} {1048} (\bibinfo {year}
  {1968})}\BibitemShut {NoStop}%
\bibitem [{\citenamefont {Cooper}\ and\ \citenamefont
  {Ringler}(1969)}]{Cooper1969}%
  \BibitemOpen
  \bibfield  {author} {\bibinfo {author} {\bibfnamefont {W.~S.}\ \bibnamefont
  {Cooper}}\ and\ \bibinfo {author} {\bibfnamefont {H.}~\bibnamefont
  {Ringler}},\ }\href {\doibase 10.1103/PhysRev.179.226} {\bibfield  {journal}
  {\bibinfo  {journal} {Phys. Rev.}\ }\textbf {\bibinfo {volume} {179}},\
  \bibinfo {pages} {226} (\bibinfo {year} {1969})}\BibitemShut {NoStop}%
\bibitem [{\citenamefont {Sprangle}\ \emph
  {et~al.}(1990{\natexlab{a}})\citenamefont {Sprangle}, \citenamefont
  {Esarey},\ and\ \citenamefont {Ting}}]{Sprangle1990a}%
  \BibitemOpen
  \bibfield  {author} {\bibinfo {author} {\bibfnamefont {P.}~\bibnamefont
  {Sprangle}}, \bibinfo {author} {\bibfnamefont {E.}~\bibnamefont {Esarey}}, \
  and\ \bibinfo {author} {\bibfnamefont {A.}~\bibnamefont {Ting}},\ }\href
  {\doibase 10.1103/PhysRevLett.64.2011} {\bibfield  {journal} {\bibinfo
  {journal} {Phys. Rev. Lett.}\ }\textbf {\bibinfo {volume} {64}},\ \bibinfo
  {pages} {2011} (\bibinfo {year} {1990}{\natexlab{a}})}\BibitemShut {NoStop}%
\bibitem [{\citenamefont {Sprangle}\ \emph
  {et~al.}(1990{\natexlab{b}})\citenamefont {Sprangle}, \citenamefont
  {Esarey},\ and\ \citenamefont {Ting}}]{Sprangle1990b}%
  \BibitemOpen
  \bibfield  {author} {\bibinfo {author} {\bibfnamefont {P.}~\bibnamefont
  {Sprangle}}, \bibinfo {author} {\bibfnamefont {E.}~\bibnamefont {Esarey}}, \
  and\ \bibinfo {author} {\bibfnamefont {A.}~\bibnamefont {Ting}},\ }\href
  {\doibase 10.1103/PhysRevA.41.4463} {\bibfield  {journal} {\bibinfo
  {journal} {Phys. Rev. A}\ }\textbf {\bibinfo {volume} {41}},\ \bibinfo
  {pages} {4463} (\bibinfo {year} {1990}{\natexlab{b}})}\BibitemShut {NoStop}%
\bibitem [{\citenamefont {Sprangle}\ \emph {et~al.}(1992)\citenamefont
  {Sprangle}, \citenamefont {Esarey}, \citenamefont {Krall},\ and\
  \citenamefont {Joyce}}]{Sprangle1992}%
  \BibitemOpen
  \bibfield  {author} {\bibinfo {author} {\bibfnamefont {P.}~\bibnamefont
  {Sprangle}}, \bibinfo {author} {\bibfnamefont {E.}~\bibnamefont {Esarey}},
  \bibinfo {author} {\bibfnamefont {J.}~\bibnamefont {Krall}}, \ and\ \bibinfo
  {author} {\bibfnamefont {G.}~\bibnamefont {Joyce}},\ }\href {\doibase
  10.1103/PhysRevLett.69.2200} {\bibfield  {journal} {\bibinfo  {journal}
  {Phys. Rev. Lett.}\ }\textbf {\bibinfo {volume} {69}},\ \bibinfo {pages}
  {2200} (\bibinfo {year} {1992})}\BibitemShut {NoStop}%
\bibitem [{\citenamefont {Esarey}\ \emph {et~al.}(1994)\citenamefont {Esarey},
  \citenamefont {Krall},\ and\ \citenamefont {Sprangle}}]{Esarey1994}%
  \BibitemOpen
  \bibfield  {author} {\bibinfo {author} {\bibfnamefont {E.}~\bibnamefont
  {Esarey}}, \bibinfo {author} {\bibfnamefont {J.}~\bibnamefont {Krall}}, \
  and\ \bibinfo {author} {\bibfnamefont {P.}~\bibnamefont {Sprangle}},\ }\href
  {\doibase 10.1103/PhysRevLett.72.2887} {\bibfield  {journal} {\bibinfo
  {journal} {Phys. Rev. Lett.}\ }\textbf {\bibinfo {volume} {72}},\ \bibinfo
  {pages} {2887} (\bibinfo {year} {1994})}\BibitemShut {NoStop}%
\bibitem [{\citenamefont {Leemans}\ \emph {et~al.}(1992)\citenamefont
  {Leemans}, \citenamefont {Clayton}, \citenamefont {Mori}, \citenamefont
  {Marsh}, \citenamefont {Kaw}, \citenamefont {Dyson}, \citenamefont {Joshi},\
  and\ \citenamefont {Wallace}}]{Leemans1992a}%
  \BibitemOpen
  \bibfield  {author} {\bibinfo {author} {\bibfnamefont {W.~P.}\ \bibnamefont
  {Leemans}}, \bibinfo {author} {\bibfnamefont {C.~E.}\ \bibnamefont
  {Clayton}}, \bibinfo {author} {\bibfnamefont {W.~B.}\ \bibnamefont {Mori}},
  \bibinfo {author} {\bibfnamefont {K.~A.}\ \bibnamefont {Marsh}}, \bibinfo
  {author} {\bibfnamefont {P.~K.}\ \bibnamefont {Kaw}}, \bibinfo {author}
  {\bibfnamefont {A.}~\bibnamefont {Dyson}}, \bibinfo {author} {\bibfnamefont
  {C.}~\bibnamefont {Joshi}}, \ and\ \bibinfo {author} {\bibfnamefont {J.~M.}\
  \bibnamefont {Wallace}},\ }\href {\doibase 10.1103/PhysRevA.46.1091}
  {\bibfield  {journal} {\bibinfo  {journal} {Phys. Rev. A}\ }\textbf {\bibinfo
  {volume} {46}},\ \bibinfo {pages} {1091} (\bibinfo {year}
  {1992})}\BibitemShut {NoStop}%
\bibitem [{\citenamefont {Mackinnon}\ \emph {et~al.}(1996)\citenamefont
  {Mackinnon}, \citenamefont {Borghesi}, \citenamefont {Iwase}, \citenamefont
  {Jones}, \citenamefont {Pert}, \citenamefont {Rae}, \citenamefont {Burnett},\
  and\ \citenamefont {Willi}}]{Mackinnon1996}%
  \BibitemOpen
  \bibfield  {author} {\bibinfo {author} {\bibfnamefont {A.~J.}\ \bibnamefont
  {Mackinnon}}, \bibinfo {author} {\bibfnamefont {M.}~\bibnamefont {Borghesi}},
  \bibinfo {author} {\bibfnamefont {A.}~\bibnamefont {Iwase}}, \bibinfo
  {author} {\bibfnamefont {M.~W.}\ \bibnamefont {Jones}}, \bibinfo {author}
  {\bibfnamefont {G.~J.}\ \bibnamefont {Pert}}, \bibinfo {author}
  {\bibfnamefont {S.}~\bibnamefont {Rae}}, \bibinfo {author} {\bibfnamefont
  {K.}~\bibnamefont {Burnett}}, \ and\ \bibinfo {author} {\bibfnamefont
  {O.}~\bibnamefont {Willi}},\ }\href {\doibase 10.1103/PhysRevLett.76.1473}
  {\bibfield  {journal} {\bibinfo  {journal} {Phys. Rev. Lett.}\ }\textbf
  {\bibinfo {volume} {76}},\ \bibinfo {pages} {1473} (\bibinfo {year}
  {1996})}\BibitemShut {NoStop}%
\bibitem [{\citenamefont {Esarey}\ \emph {et~al.}(1997)\citenamefont {Esarey},
  \citenamefont {Sprangle}, \citenamefont {Krall},\ and\ \citenamefont
  {Ting}}]{Esarey1997}%
  \BibitemOpen
  \bibfield  {author} {\bibinfo {author} {\bibfnamefont {E.}~\bibnamefont
  {Esarey}}, \bibinfo {author} {\bibfnamefont {P.}~\bibnamefont {Sprangle}},
  \bibinfo {author} {\bibfnamefont {J.}~\bibnamefont {Krall}}, \ and\ \bibinfo
  {author} {\bibfnamefont {A.}~\bibnamefont {Ting}},\ }\href {\doibase
  10.1109/3.641305} {\bibfield  {journal} {\bibinfo  {journal} {IEEE J. Quant.
  Elec.}\ }\textbf {\bibinfo {volume} {33}},\ \bibinfo {pages} {1879 }
  (\bibinfo {year} {1997})}\BibitemShut {NoStop}%
\bibitem [{\citenamefont {Sprangle}\ \emph {et~al.}(1996)\citenamefont
  {Sprangle}, \citenamefont {Esarey},\ and\ \citenamefont
  {Krall}}]{Sprangle1996}%
  \BibitemOpen
  \bibfield  {author} {\bibinfo {author} {\bibfnamefont {P.}~\bibnamefont
  {Sprangle}}, \bibinfo {author} {\bibfnamefont {E.}~\bibnamefont {Esarey}}, \
  and\ \bibinfo {author} {\bibfnamefont {J.}~\bibnamefont {Krall}},\ }\href
  {\doibase 10.1103/PhysRevE.54.4211} {\bibfield  {journal} {\bibinfo
  {journal} {Phys. Rev. E}\ }\textbf {\bibinfo {volume} {54}},\ \bibinfo
  {pages} {4211} (\bibinfo {year} {1996})}\BibitemShut {NoStop}%
\bibitem [{\citenamefont {Wu}\ and\ \citenamefont {Jr}(2003)}]{Wu2003}%
  \BibitemOpen
  \bibfield  {author} {\bibinfo {author} {\bibfnamefont {J.}~\bibnamefont
  {Wu}}\ and\ \bibinfo {author} {\bibfnamefont {T.~M.~A.}\ \bibnamefont {Jr}},\
  }\href {\doibase 10.1063/1.1571544} {\bibfield  {journal} {\bibinfo
  {journal} {Phys. Plasmas}\ }\textbf {\bibinfo {volume} {10}},\ \bibinfo
  {pages} {2254} (\bibinfo {year} {2003})}\BibitemShut {NoStop}%
\bibitem [{\citenamefont {Esarey}\ \emph {et~al.}(2009)\citenamefont {Esarey},
  \citenamefont {Schroeder},\ and\ \citenamefont
  {Leemans}}]{RevModPhys.81.1229}%
  \BibitemOpen
  \bibfield  {author} {\bibinfo {author} {\bibfnamefont {E.}~\bibnamefont
  {Esarey}}, \bibinfo {author} {\bibfnamefont {C.~B.}\ \bibnamefont
  {Schroeder}}, \ and\ \bibinfo {author} {\bibfnamefont {W.~P.}\ \bibnamefont
  {Leemans}},\ }\href {\doibase 10.1103/RevModPhys.81.1229} {\bibfield
  {journal} {\bibinfo  {journal} {Rev. Mod. Phys.}\ }\textbf {\bibinfo {volume}
  {81}},\ \bibinfo {pages} {1229} (\bibinfo {year} {2009})}\BibitemShut
  {NoStop}%
\bibitem [{\citenamefont {Hu}\ \emph {et~al.}(2005)\citenamefont {Hu},
  \citenamefont {Liu}, \citenamefont {Jiang},\ and\ \citenamefont
  {Zhang}}]{Hu2005}%
  \BibitemOpen
  \bibfield  {author} {\bibinfo {author} {\bibfnamefont {Q.-L.}\ \bibnamefont
  {Hu}}, \bibinfo {author} {\bibfnamefont {S.-B.}\ \bibnamefont {Liu}},
  \bibinfo {author} {\bibfnamefont {Y.~J.}\ \bibnamefont {Jiang}}, \ and\
  \bibinfo {author} {\bibfnamefont {J.}~\bibnamefont {Zhang}},\ }\href
  {\doibase doi:10.1063/1.2007887} {\bibfield  {journal} {\bibinfo  {journal}
  {Phys. Plasmas}\ }\textbf {\bibinfo {volume} {12}},\ \bibinfo {pages}
  {083104} (\bibinfo {year} {2005})}\BibitemShut {NoStop}%
\bibitem [{\citenamefont {Xia}\ \emph {et~al.}(2009)\citenamefont {Xia},
  \citenamefont {Yin},\ and\ \citenamefont {Kresin}}]{PhysRevLett.102.156802}%
  \BibitemOpen
  \bibfield  {author} {\bibinfo {author} {\bibfnamefont {C.}~\bibnamefont
  {Xia}}, \bibinfo {author} {\bibfnamefont {C.}~\bibnamefont {Yin}}, \ and\
  \bibinfo {author} {\bibfnamefont {V.~V.}\ \bibnamefont {Kresin}},\ }\href
  {\doibase 10.1103/PhysRevLett.102.156802} {\bibfield  {journal} {\bibinfo
  {journal} {Phys. Rev. Lett.}\ }\textbf {\bibinfo {volume} {102}},\ \bibinfo
  {pages} {156802} (\bibinfo {year} {2009})}\BibitemShut {NoStop}%
\bibitem [{\citenamefont {Lombardi}\ and\ \citenamefont
  {Birke}(2012)}]{:/content/aip/journal/jcp/136/14/10.1063/1.3698292}%
  \BibitemOpen
  \bibfield  {author} {\bibinfo {author} {\bibfnamefont {J.~R.}\ \bibnamefont
  {Lombardi}}\ and\ \bibinfo {author} {\bibfnamefont {R.~L.}\ \bibnamefont
  {Birke}},\ }\href {\doibase http://dx.doi.org/10.1063/1.3698292} {\bibfield
  {journal} {\bibinfo  {journal} {J. Chem. Phys.}\ }\textbf {\bibinfo {volume}
  {136}},\ \bibinfo {eid} {144704} (\bibinfo {year} {2012})}\BibitemShut
  {NoStop}%
\bibitem [{\citenamefont {Sitenko}\ and\ \citenamefont
  {Malnev}(1995)}]{SitenkoMalnev1995}%
  \BibitemOpen
  \bibfield  {author} {\bibinfo {author} {\bibfnamefont {A.}~\bibnamefont
  {Sitenko}}\ and\ \bibinfo {author} {\bibfnamefont {V.}~\bibnamefont
  {Malnev}},\ }\href
  {http://books.google.ca/books?id=NhKWNZIohRwC&lpg=PP1&ots=o0SkslFwLt&dq=sitenko%20plasma&pg=PP1#v=onepage&q&f=false}
  {\emph {\bibinfo {title} {Plasma Physics Theory}}}\ (\bibinfo  {publisher}
  {Chapman \& Hall, London},\ \bibinfo {year} {1995})\BibitemShut {NoStop}%
\bibitem [{\citenamefont {Liboff}(2003)}]{Liboff.R.L2003}%
  \BibitemOpen
  \bibfield  {author} {\bibinfo {author} {\bibfnamefont {R.~L.}\ \bibnamefont
  {Liboff}},\ }\href@noop {} {\emph {\bibinfo {title} {Kinetic Theory:
  Classical, Quantum, and Relativistic Descriptions, Third Ed.}}}\ (\bibinfo
  {publisher} {Springer},\ \bibinfo {year} {2003})\BibitemShut {NoStop}%
\bibitem [{\citenamefont {Goldston}\ and\ \citenamefont
  {Rutherford}(1995)}]{GoldstonRutherford}%
  \BibitemOpen
  \bibfield  {author} {\bibinfo {author} {\bibfnamefont {R.}~\bibnamefont
  {Goldston}}\ and\ \bibinfo {author} {\bibfnamefont {P.}~\bibnamefont
  {Rutherford}},\ }\href@noop {} {\emph {\bibinfo {title} {Introduction to
  Plasma Physics}}}\ (\bibinfo  {publisher} {CRC Press},\ \bibinfo {year}
  {1995})\BibitemShut {NoStop}%
\bibitem [{\citenamefont {Lombardi}\ and\ \citenamefont
  {Birke}(2008)}]{doi:10.1021/jp800167v}%
  \BibitemOpen
  \bibfield  {author} {\bibinfo {author} {\bibfnamefont {J.~R.}\ \bibnamefont
  {Lombardi}}\ and\ \bibinfo {author} {\bibfnamefont {R.~L.}\ \bibnamefont
  {Birke}},\ }\href {\doibase 10.1021/jp800167v} {\bibfield  {journal}
  {\bibinfo  {journal} {J. Phys. Chem. C}\ }\textbf {\bibinfo {volume} {112}},\
  \bibinfo {pages} {5605} (\bibinfo {year} {2008})}\BibitemShut {NoStop}%
\bibitem [{\citenamefont {Ferguson}\ \emph {et~al.}(2002)\citenamefont
  {Ferguson}, \citenamefont {Cain}, \citenamefont {Williams},\ and\
  \citenamefont {Briggs}}]{Ferguson2002}%
  \BibitemOpen
  \bibfield  {author} {\bibinfo {author} {\bibfnamefont {A.~J.}\ \bibnamefont
  {Ferguson}}, \bibinfo {author} {\bibfnamefont {P.~A.}\ \bibnamefont {Cain}},
  \bibinfo {author} {\bibfnamefont {D.~A.}\ \bibnamefont {Williams}}, \ and\
  \bibinfo {author} {\bibfnamefont {G.~A.~D.}\ \bibnamefont {Briggs}},\ }\href
  {\doibase 10.1103/PhysRevA.65.034303} {\bibfield  {journal} {\bibinfo
  {journal} {Phys. Rev. A}\ }\textbf {\bibinfo {volume} {65}},\ \bibinfo
  {pages} {034303} (\bibinfo {year} {2002})}\BibitemShut {NoStop}%
\bibitem [{\citenamefont {Schr{\"o}der}\ and\ \citenamefont
  {Brown}(2009)}]{Schroder2009}%
  \BibitemOpen
  \bibfield  {author} {\bibinfo {author} {\bibfnamefont {M.}~\bibnamefont
  {Schr{\"o}der}}\ and\ \bibinfo {author} {\bibfnamefont {A.}~\bibnamefont
  {Brown}},\ }\href {\doibase doi:10.1063/1.3168438} {\bibfield  {journal}
  {\bibinfo  {journal} {J. Chem. Phys.}\ }\textbf {\bibinfo {volume} {131}},\
  \bibinfo {pages} {034101} (\bibinfo {year} {2009})}\BibitemShut {NoStop}%
\bibitem [{\citenamefont {Soderberg}\ \emph {et~al.}(2009)\citenamefont
  {Soderberg}, \citenamefont {Gemelke},\ and\ \citenamefont
  {Chin}}]{1367-2630-11-5-055022}%
  \BibitemOpen
  \bibfield  {author} {\bibinfo {author} {\bibfnamefont {K.-A.~B.}\
  \bibnamefont {Soderberg}}, \bibinfo {author} {\bibfnamefont {N.}~\bibnamefont
  {Gemelke}}, \ and\ \bibinfo {author} {\bibfnamefont {C.}~\bibnamefont
  {Chin}},\ }\href {http://stacks.iop.org/1367-2630/11/i=5/a=055022} {\bibfield
   {journal} {\bibinfo  {journal} {New J. Phys.}\ }\textbf {\bibinfo {volume}
  {11}},\ \bibinfo {pages} {055022} (\bibinfo {year} {2009})}\BibitemShut
  {NoStop}%
\bibitem [{\citenamefont {Cohen-Tannoudji}(1992)}]{TannoudjiLesHouches1990}%
  \BibitemOpen
  \bibfield  {author} {\bibinfo {author} {\bibfnamefont {C.}~\bibnamefont
  {Cohen-Tannoudji}},\ }in\ \href@noop {} {\emph {\bibinfo {booktitle}
  {Fundamental Systems in Quantum Optics}}},\ \bibinfo {series} {Les Houches},
  Vol.\ \bibinfo {volume} {LIII (1990)},\ \bibinfo {editor} {edited by\
  \bibinfo {editor} {\bibfnamefont {J.}~\bibnamefont {Dalibard}}, \bibinfo
  {editor} {\bibfnamefont {J.}~\bibnamefont {Raimond}}, \ and\ \bibinfo
  {editor} {\bibfnamefont {J.~Z.}\ \bibnamefont {Justin}}}\ (\bibinfo
  {publisher} {Elsevier Science Publisher B.V.},\ \bibinfo {year} {1992})\ pp.\
  \bibinfo {pages} {1--164}\BibitemShut {NoStop}%
\bibitem [{\citenamefont {Cohen-Tannoudji}\ \emph {et~al.}(2008)\citenamefont
  {Cohen-Tannoudji}, \citenamefont {Dupont-Roc},\ and\ \citenamefont
  {Grynberg}}]{AtomPhotonInteractions}%
  \BibitemOpen
  \bibfield  {author} {\bibinfo {author} {\bibfnamefont {C.}~\bibnamefont
  {Cohen-Tannoudji}}, \bibinfo {author} {\bibfnamefont {J.}~\bibnamefont
  {Dupont-Roc}}, \ and\ \bibinfo {author} {\bibfnamefont {G.}~\bibnamefont
  {Grynberg}},\ }\href {\doibase 10.1002/9783527617197} {\emph {\bibinfo
  {title} {Atom-Photon Interactions: Basic Processes and Appilcations}}}\
  (\bibinfo  {publisher} {WILEY-VCH, Weinheim, Germany},\ \bibinfo {year}
  {2008})\BibitemShut {NoStop}%
\bibitem [{\citenamefont {Kneipp}\ \emph {et~al.}(2006)\citenamefont {Kneipp},
  \citenamefont {Moskovits},\ and\ \citenamefont {Kneipp}}]{KneippSERS2006}%
  \BibitemOpen
  \bibinfo {editor} {\bibfnamefont {K.}~\bibnamefont {Kneipp}}, \bibinfo
  {editor} {\bibfnamefont {M.}~\bibnamefont {Moskovits}}, \ and\ \bibinfo
  {editor} {\bibfnamefont {H.}~\bibnamefont {Kneipp}},\ eds.,\ \href@noop {}
  {\emph {\bibinfo {title} {Surface-enhanced Raman scattering}}}\ (\bibinfo
  {publisher} {Springer},\ \bibinfo {year} {2006})\BibitemShut {NoStop}%
\bibitem [{\citenamefont {Fleischhauer}\ and\ \citenamefont
  {Yelin}(1999)}]{PRA59:2427}%
  \BibitemOpen
  \bibfield  {author} {\bibinfo {author} {\bibfnamefont {M.}~\bibnamefont
  {Fleischhauer}}\ and\ \bibinfo {author} {\bibfnamefont {S.~F.}\ \bibnamefont
  {Yelin}},\ }\href {http://link.aps.org/abstract/PRA/v59/p2427} {\bibfield
  {journal} {\bibinfo  {journal} {Phys. Rev. A}\ }\textbf {\bibinfo {volume}
  {59}},\ \bibinfo {pages} {2427} (\bibinfo {year} {1999})}\BibitemShut
  {NoStop}%
\bibitem [{\citenamefont {Motsch}\ \emph {et~al.}(2009)\citenamefont {Motsch},
  \citenamefont {Sommer}, \citenamefont {Zeppenfeld}, \citenamefont {van
  Buuren}, \citenamefont {Pinkse},\ and\ \citenamefont
  {Rempe}}]{1367-2630-11-5-055030}%
  \BibitemOpen
  \bibfield  {author} {\bibinfo {author} {\bibfnamefont {M.}~\bibnamefont
  {Motsch}}, \bibinfo {author} {\bibfnamefont {C.}~\bibnamefont {Sommer}},
  \bibinfo {author} {\bibfnamefont {M.}~\bibnamefont {Zeppenfeld}}, \bibinfo
  {author} {\bibfnamefont {L.~D.}\ \bibnamefont {van Buuren}}, \bibinfo
  {author} {\bibfnamefont {P.~W.~H.}\ \bibnamefont {Pinkse}}, \ and\ \bibinfo
  {author} {\bibfnamefont {G.}~\bibnamefont {Rempe}},\ }\href
  {http://stacks.iop.org/1367-2630/11/i=5/a=055030} {\bibfield  {journal}
  {\bibinfo  {journal} {New J. Phys.}\ }\textbf {\bibinfo {volume} {11}},\
  \bibinfo {pages} {055030} (\bibinfo {year} {2009})}\BibitemShut {NoStop}%
\bibitem [{\citenamefont {Marquardt}\ \emph {et~al.}(2003)\citenamefont
  {Marquardt}, \citenamefont {Quack}, \citenamefont {Thanopulos},\ and\
  \citenamefont {Luckhaus}}]{10.1063/1.1617272}%
  \BibitemOpen
  \bibfield  {author} {\bibinfo {author} {\bibfnamefont {R.}~\bibnamefont
  {Marquardt}}, \bibinfo {author} {\bibfnamefont {M.}~\bibnamefont {Quack}},
  \bibinfo {author} {\bibfnamefont {I.}~\bibnamefont {Thanopulos}}, \ and\
  \bibinfo {author} {\bibfnamefont {D.}~\bibnamefont {Luckhaus}},\ }\href
  {\doibase DOI:10.1063/1.1617272} {\bibfield  {journal} {\bibinfo  {journal}
  {J. Chem. Phys.}\ }\textbf {\bibinfo {volume} {119}},\ \bibinfo {pages}
  {10724} (\bibinfo {year} {2003})}\BibitemShut {NoStop}%
\bibitem [{\citenamefont {Moiseev}\ and\ \citenamefont
  {Kr\"oll}(2001)}]{PhysRevLett.87.173601}%
  \BibitemOpen
  \bibfield  {author} {\bibinfo {author} {\bibfnamefont {S.~A.}\ \bibnamefont
  {Moiseev}}\ and\ \bibinfo {author} {\bibfnamefont {S.}~\bibnamefont
  {Kr\"oll}},\ }\href {\doibase 10.1103/PhysRevLett.87.173601} {\bibfield
  {journal} {\bibinfo  {journal} {Phys. Rev. Lett.}\ }\textbf {\bibinfo
  {volume} {87}},\ \bibinfo {pages} {173601} (\bibinfo {year}
  {2001})}\BibitemShut {NoStop}%
\bibitem [{\citenamefont {Fleischhauer}\ and\ \citenamefont
  {Lukin}(2002)}]{PRA65:022314}%
  \BibitemOpen
  \bibfield  {author} {\bibinfo {author} {\bibfnamefont {M.}~\bibnamefont
  {Fleischhauer}}\ and\ \bibinfo {author} {\bibfnamefont {M.~D.}\ \bibnamefont
  {Lukin}},\ }\href {http://link.aps.org/abstract/PRA/v65/e022314} {\bibfield
  {journal} {\bibinfo  {journal} {Phys. Rev. A}\ }\textbf {\bibinfo {volume}
  {65}},\ \bibinfo {pages} {022314} (\bibinfo {year} {2002})}\BibitemShut
  {NoStop}%
\bibitem [{\citenamefont {Zhang}\ and\ \citenamefont
  {Walls}(1994)}]{PhysRevA.49.3799}%
  \BibitemOpen
  \bibfield  {author} {\bibinfo {author} {\bibfnamefont {W.}~\bibnamefont
  {Zhang}}\ and\ \bibinfo {author} {\bibfnamefont {D.~F.}\ \bibnamefont
  {Walls}},\ }\href {\doibase 10.1103/PhysRevA.49.3799} {\bibfield  {journal}
  {\bibinfo  {journal} {Phys. Rev. A}\ }\textbf {\bibinfo {volume} {49}},\
  \bibinfo {pages} {3799} (\bibinfo {year} {1994})}\BibitemShut {NoStop}%
\bibitem [{\citenamefont {Lenz}\ \emph {et~al.}(1994)\citenamefont {Lenz},
  \citenamefont {Meystre},\ and\ \citenamefont {Wright}}]{PhysRevA.50.1681}%
  \BibitemOpen
  \bibfield  {author} {\bibinfo {author} {\bibfnamefont {G.}~\bibnamefont
  {Lenz}}, \bibinfo {author} {\bibfnamefont {P.}~\bibnamefont {Meystre}}, \
  and\ \bibinfo {author} {\bibfnamefont {E.~M.}\ \bibnamefont {Wright}},\
  }\href {\doibase 10.1103/PhysRevA.50.1681} {\bibfield  {journal} {\bibinfo
  {journal} {Phys. Rev. A}\ }\textbf {\bibinfo {volume} {50}},\ \bibinfo
  {pages} {1681} (\bibinfo {year} {1994})}\BibitemShut {NoStop}%
\bibitem [{\citenamefont {Jackson}(1998)}]{JacksonEdyn}%
  \BibitemOpen
  \bibfield  {author} {\bibinfo {author} {\bibfnamefont {J.~D.}\ \bibnamefont
  {Jackson}},\ }\href@noop {} {\emph {\bibinfo {title} {Classical
  Electrodynamics}}},\ \bibinfo {edition} {3rd}\ ed.\ (\bibinfo  {publisher}
  {Wiley},\ \bibinfo {address} {New York},\ \bibinfo {year} {1998})\BibitemShut
  {NoStop}%
\end{thebibliography}%
\bibliographystyle{/Users/pmarzlin/tex/revtex4-1/revtex4-1-tds/bibtex/bst/revtex/apsrev4-1}

\end{document}